\documentclass[longauth]{aa}

\usepackage{graphicx}
\usepackage{natbib}
\usepackage{scalerel}
\usepackage{lastpage}

\usepackage[table]{xcolor}

\usepackage{txfonts}
\usepackage[pdfencoding=auto,psdextra]{hyperref}
\hypersetup{
    colorlinks=true,
    linkcolor=blue,
    filecolor=magenta,      
    urlcolor=blue,
    citecolor=blue
}
\makeatletter
\renewcommand*\aa@pageof{, page \thepage{} of \pageref*{LastPage}}
\makeatother

\usepackage[utf8]{inputenc}

\usepackage[switch, modulo]{lineno}
              
\usepackage{euclid}

\def\Sref#1{Sect.~\ref{#1}\xspace}
\def\Fref#1{Fig.\,\ref{#1}\xspace}
\def\Tref#1{Table~\ref{#1}\xspace}
\def\Eref#1{Eq.~(\ref{#1})\xspace}

\newcommand{\mujansky}{\mu\mathrm{Jy}}

\def\Ngold{95}

\def\NSLDE{80}
\def\Namb{17}
\def\Nmodelled{408}
\def\Nmanual{40}
\def\Nspecz{18}

\def\fluxvis{F_{\mathrm{g},\IE}}
\def\fluxset{\{F_{\mathrm{g},\lambda}\}}

\def\tein{\theta_\sfont{E}}
\def\teinref{\theta_\sfont{E}^*}
\def\teff{\theta_{\mathrm{e}}}
\def\mein{M_\sfont{E}}
\def\mfive{M_5}
\def\fivekpc{5\,h^{-1}\,\mathrm{kpc}}
\def\reff{R_{\mathrm{e}}}

\def\SN{\mathrm{S}/\mathrm{N}}

\def\zlens{z_\mathrm{g}}
\def\zsource{z_\mathrm{s}}
\def\zt{z_\mathrm{t}}

\def\lensflux{F_{\mathrm{g},\IE}}
\def\sourceflux{F_{\mathrm{s},\IE}}

\def\azpar{a_z}
\def\czpar{c_z}
\def\zscalepar{z_0}

\newcommand{\prlens}{P_{\mathrm{g}}}
\newcommand{\prstar}{P_*}
\newcommand{\prmass}{P_{\mathrm{M}}}
\newcommand{\prsource}{P_{\mathrm{s}}}
\newcommand{\prsl}{P_{\mathrm{SL}}}
\newcommand{\psel}{P_{\mathrm{sel}}}
\newcommand{\crosssect}{\sigma_{\mathrm{SL}}}

\def\data{\vec{d}}
\def\datai{\vec{d}_i}

\def\hyperpars{\vec{\eta}}

\def\hyperparsref{\vec\eta^*}

\def\psilens{\vec{\psi}_\mathrm{g}}

\def\psilensi{\vec{\psi}_{\mathrm{g},i}}
\def\psisource{\vec{\psi}_\mathrm{s}}

\def\psisourcei{\vec{\psi}_{\mathrm{s},i}}
\def\psisourcenobeta{\vec{\psi}_\mathrm{s}^{(-\vec\beta)}}
\def\sourcepos{\vec\beta}

\defcitealias{SLDEA}{W26}
\newcommand{\SLDEA}{\citetalias{SLDEA}}

\begin{document}
%


\title{Euclid Quick Data Release (Q1)} 
 \subtitle{Mass-to-light ratio of elliptical galaxies: constraints from strong lensing}
%

   
\newcommand{\orcid}[1]{} 
\author{Euclid Collaboration: A.~Sonnenfeld\orcid{0000-0002-6061-5977}\thanks{sonnenfeld@sjtu.edu.cn}\inst{\ref{aff1},\ref{aff2},\ref{aff3}}
\and A.~Nersesian\orcid{0000-0001-6843-409X}\inst{\ref{aff4},\ref{aff5}}
\and K.~Rojas\orcid{0000-0003-1391-6854}\inst{\ref{aff6}}
\and P.~Holloway\orcid{0009-0002-8896-6100}\inst{\ref{aff7}}
\and J.~W.~Nightingale\orcid{0000-0002-8987-7401}\inst{\ref{aff8}}
\and C.~Tortora\orcid{0000-0001-7958-6531}\inst{\ref{aff9}}
\and S.~Schuldt\orcid{0000-0003-2497-6334}\inst{\ref{aff10},\ref{aff11},\ref{aff12}}
\and T.~Li\orcid{0009-0005-5008-0381}\inst{\ref{aff7}}
\and C.~J.~Storfer\orcid{0000-0002-0385-0014}\inst{\ref{aff13}}
\and R.~Gavazzi\orcid{0000-0002-5540-6935}\inst{\ref{aff14},\ref{aff15}}
\and D.~Sluse\orcid{0000-0001-6116-2095}\inst{\ref{aff4}}
\and G.~Despali\orcid{0000-0001-6150-4112}\inst{\ref{aff16},\ref{aff17},\ref{aff18}}
\and T.~E.~Collett\orcid{0000-0001-5564-3140}\inst{\ref{aff7}}
\and S.~Serjeant\orcid{0000-0002-0517-7943}\inst{\ref{aff19}}
\and F.~Courbin\orcid{0000-0003-0758-6510}\inst{\ref{aff20},\ref{aff21},\ref{aff22}}
\and R.~B.~Metcalf\orcid{0000-0003-3167-2574}\inst{\ref{aff16},\ref{aff17}}
\and S.~Vegetti\orcid{0009-0006-0592-2882}\inst{\ref{aff23}}
\and B.~Altieri\orcid{0000-0003-3936-0284}\inst{\ref{aff24}}
\and S.~Andreon\orcid{0000-0002-2041-8784}\inst{\ref{aff25}}
\and N.~Auricchio\orcid{0000-0003-4444-8651}\inst{\ref{aff17}}
\and C.~Baccigalupi\orcid{0000-0002-8211-1630}\inst{\ref{aff26},\ref{aff27},\ref{aff28},\ref{aff29}}
\and M.~Baldi\orcid{0000-0003-4145-1943}\inst{\ref{aff30},\ref{aff17},\ref{aff18}}
\and A.~Balestra\orcid{0000-0002-6967-261X}\inst{\ref{aff31}}
\and S.~Bardelli\orcid{0000-0002-8900-0298}\inst{\ref{aff17}}
\and P.~Battaglia\orcid{0000-0002-7337-5909}\inst{\ref{aff17}}
\and A.~Biviano\orcid{0000-0002-0857-0732}\inst{\ref{aff27},\ref{aff26}}
\and M.~Bolzonella\orcid{0000-0003-3278-4607}\inst{\ref{aff17}}
\and E.~Branchini\orcid{0000-0002-0808-6908}\inst{\ref{aff32},\ref{aff33},\ref{aff25}}
\and M.~Brescia\orcid{0000-0001-9506-5680}\inst{\ref{aff34},\ref{aff9}}
\and J.~Brinchmann\orcid{0000-0003-4359-8797}\inst{\ref{aff35},\ref{aff36},\ref{aff37}}
\and S.~Camera\orcid{0000-0003-3399-3574}\inst{\ref{aff38},\ref{aff39},\ref{aff40}}
\and G.~Ca\~nas-Herrera\orcid{0000-0003-2796-2149}\inst{\ref{aff41}}
\and V.~Capobianco\orcid{0000-0002-3309-7692}\inst{\ref{aff40}}
\and C.~Carbone\orcid{0000-0003-0125-3563}\inst{\ref{aff12}}
\and M.~Castellano\orcid{0000-0001-9875-8263}\inst{\ref{aff42}}
\and G.~Castignani\orcid{0000-0001-6831-0687}\inst{\ref{aff17}}
\and S.~Cavuoti\orcid{0000-0002-3787-4196}\inst{\ref{aff9},\ref{aff43}}
\and K.~C.~Chambers\orcid{0000-0001-6965-7789}\inst{\ref{aff13}}
\and A.~Cimatti\inst{\ref{aff44}}
\and C.~Colodro-Conde\inst{\ref{aff45}}
\and G.~Congedo\orcid{0000-0003-2508-0046}\inst{\ref{aff46}}
\and C.~J.~Conselice\orcid{0000-0003-1949-7638}\inst{\ref{aff47}}
\and L.~Conversi\orcid{0000-0002-6710-8476}\inst{\ref{aff48},\ref{aff24}}
\and Y.~Copin\orcid{0000-0002-5317-7518}\inst{\ref{aff49}}
\and A.~Costille\inst{\ref{aff14}}
\and H.~M.~Courtois\orcid{0000-0003-0509-1776}\inst{\ref{aff50}}
\and M.~Cropper\orcid{0000-0003-4571-9468}\inst{\ref{aff51}}
\and H.~Degaudenzi\orcid{0000-0002-5887-6799}\inst{\ref{aff52}}
\and G.~De~Lucia\orcid{0000-0002-6220-9104}\inst{\ref{aff27}}
\and C.~Dolding\orcid{0009-0003-7199-6108}\inst{\ref{aff51}}
\and H.~Dole\orcid{0000-0002-9767-3839}\inst{\ref{aff53}}
\and F.~Dubath\orcid{0000-0002-6533-2810}\inst{\ref{aff52}}
\and C.~A.~J.~Duncan\orcid{0009-0003-3573-0791}\inst{\ref{aff46}}
\and X.~Dupac\inst{\ref{aff24}}
\and S.~Dusini\orcid{0000-0002-1128-0664}\inst{\ref{aff54}}
\and S.~Escoffier\orcid{0000-0002-2847-7498}\inst{\ref{aff55}}
\and M.~Farina\orcid{0000-0002-3089-7846}\inst{\ref{aff56}}
\and R.~Farinelli\inst{\ref{aff17}}
\and F.~Faustini\orcid{0000-0001-6274-5145}\inst{\ref{aff42}}
\and S.~Ferriol\inst{\ref{aff49}}
\and F.~Finelli\orcid{0000-0002-6694-3269}\inst{\ref{aff17},\ref{aff57}}
\and P.~Fosalba\orcid{0000-0002-1510-5214}\inst{\ref{aff58},\ref{aff59}}
\and M.~Frailis\orcid{0000-0002-7400-2135}\inst{\ref{aff27}}
\and E.~Franceschi\orcid{0000-0002-0585-6591}\inst{\ref{aff17}}
\and M.~Fumana\orcid{0000-0001-6787-5950}\inst{\ref{aff12}}
\and L.~Gabarra\orcid{0000-0002-8486-8856}\inst{\ref{aff60}}
\and S.~Galeotta\orcid{0000-0002-3748-5115}\inst{\ref{aff27}}
\and K.~George\orcid{0000-0002-1734-8455}\inst{\ref{aff61}}
\and B.~Gillis\orcid{0000-0002-4478-1270}\inst{\ref{aff46}}
\and C.~Giocoli\orcid{0000-0002-9590-7961}\inst{\ref{aff17},\ref{aff18}}
\and J.~Gracia-Carpio\orcid{0000-0003-4689-3134}\inst{\ref{aff62}}
\and A.~Grazian\orcid{0000-0002-5688-0663}\inst{\ref{aff31}}
\and F.~Grupp\inst{\ref{aff62},\ref{aff63}}
\and W.~G.~Hartley\orcid{0000-0002-4563-5101}\inst{\ref{aff52}}
\and S.~V.~H.~Haugan\orcid{0000-0001-9648-7260}\inst{\ref{aff64}}
\and G.~Helou\orcid{0000-0003-3367-3415}\inst{\ref{aff65},\ref{aff66}}
\and S.~Hemmati\orcid{0000-0003-2226-5395}\inst{\ref{aff66}}
\and H.~Hoekstra\orcid{0000-0002-0641-3231}\inst{\ref{aff41}}
\and W.~Holmes\orcid{0009-0007-8554-4646}\inst{\ref{aff67}}
\and I.~M.~Hook\orcid{0000-0002-2960-978X}\inst{\ref{aff68}}
\and F.~Hormuth\inst{\ref{aff69}}
\and A.~Hornstrup\orcid{0000-0002-3363-0936}\inst{\ref{aff70},\ref{aff71}}
\and K.~Jahnke\orcid{0000-0003-3804-2137}\inst{\ref{aff72}}
\and M.~Jhabvala\inst{\ref{aff73}}
\and S.~Kermiche\orcid{0000-0002-0302-5735}\inst{\ref{aff55}}
\and A.~Kiessling\orcid{0000-0002-2590-1273}\inst{\ref{aff67}}
\and B.~Kubik\orcid{0009-0006-5823-4880}\inst{\ref{aff49}}
\and K.~Kuijken\orcid{0000-0002-3827-0175}\inst{\ref{aff41}}
\and M.~K\"ummel\orcid{0000-0003-2791-2117}\inst{\ref{aff63}}
\and M.~Kunz\orcid{0000-0002-3052-7394}\inst{\ref{aff74}}
\and H.~Kurki-Suonio\orcid{0000-0002-4618-3063}\inst{\ref{aff11},\ref{aff75}}
\and A.~M.~C.~Le~Brun\orcid{0000-0002-0936-4594}\inst{\ref{aff76}}
\and D.~Le~Mignant\orcid{0000-0002-5339-5515}\inst{\ref{aff14}}
\and S.~Ligori\orcid{0000-0003-4172-4606}\inst{\ref{aff40}}
\and P.~B.~Lilje\orcid{0000-0003-4324-7794}\inst{\ref{aff64}}
\and V.~Lindholm\orcid{0000-0003-2317-5471}\inst{\ref{aff11},\ref{aff75}}
\and I.~Lloro\orcid{0000-0001-5966-1434}\inst{\ref{aff77}}
\and M.~Magliocchetti\orcid{0000-0001-9158-4838}\inst{\ref{aff56}}
\and G.~Mainetti\orcid{0000-0003-2384-2377}\inst{\ref{aff78}}
\and O.~Mansutti\orcid{0000-0001-5758-4658}\inst{\ref{aff27}}
\and S.~Marcin\inst{\ref{aff6}}
\and O.~Marggraf\orcid{0000-0001-7242-3852}\inst{\ref{aff79}}
\and M.~Martinelli\orcid{0000-0002-6943-7732}\inst{\ref{aff42},\ref{aff80}}
\and N.~Martinet\orcid{0000-0003-2786-7790}\inst{\ref{aff14}}
\and F.~Marulli\orcid{0000-0002-8850-0303}\inst{\ref{aff16},\ref{aff17},\ref{aff18}}
\and R.~J.~Massey\orcid{0000-0002-6085-3780}\inst{\ref{aff81}}
\and N.~Mauri\orcid{0000-0001-8196-1548}\inst{\ref{aff44},\ref{aff18}}
\and E.~Medinaceli\orcid{0000-0002-4040-7783}\inst{\ref{aff17}}
\and S.~Mei\orcid{0000-0002-2849-559X}\inst{\ref{aff82},\ref{aff83}}
\and M.~Meneghetti\orcid{0000-0003-1225-7084}\inst{\ref{aff17},\ref{aff18}}
\and E.~Merlin\orcid{0000-0001-6870-8900}\inst{\ref{aff31}}
\and G.~Meylan\orcid{0000-0001-6503-0209}\inst{\ref{aff84}}
\and P.~Monaco\orcid{0000-0003-2083-7564}\inst{\ref{aff85},\ref{aff27},\ref{aff28},\ref{aff26}}
\and A.~Mora\orcid{0000-0002-1922-8529}\inst{\ref{aff86}}
\and M.~Moresco\orcid{0000-0002-7616-7136}\inst{\ref{aff16},\ref{aff17}}
\and L.~Moscardini\orcid{0000-0002-3473-6716}\inst{\ref{aff16},\ref{aff17},\ref{aff18}}
\and R.~Nakajima\orcid{0009-0009-1213-7040}\inst{\ref{aff79}}
\and C.~Neissner\orcid{0000-0001-8524-4968}\inst{\ref{aff87},\ref{aff88}}
\and R.~C.~Nichol\orcid{0000-0003-0939-6518}\inst{\ref{aff89}}
\and S.-M.~Niemi\orcid{0009-0005-0247-0086}\inst{\ref{aff90}}
\and C.~Padilla\orcid{0000-0001-7951-0166}\inst{\ref{aff87}}
\and S.~Paltani\orcid{0000-0002-8108-9179}\inst{\ref{aff52}}
\and F.~Pasian\orcid{0000-0002-4869-3227}\inst{\ref{aff27}}
\and W.~J.~Percival\orcid{0000-0002-0644-5727}\inst{\ref{aff91},\ref{aff92},\ref{aff93}}
\and V.~Pettorino\orcid{0000-0002-4203-9320}\inst{\ref{aff90}}
\and S.~Pires\orcid{0000-0002-0249-2104}\inst{\ref{aff94}}
\and G.~Polenta\orcid{0000-0003-4067-9196}\inst{\ref{aff95}}
\and M.~Poncet\inst{\ref{aff96}}
\and L.~A.~Popa\inst{\ref{aff97}}
\and L.~Pozzetti\orcid{0000-0001-7085-0412}\inst{\ref{aff17}}
\and F.~Raison\orcid{0000-0002-7819-6918}\inst{\ref{aff62}}
\and A.~Renzi\orcid{0000-0001-9856-1970}\inst{\ref{aff98},\ref{aff54},\ref{aff17}}
\and J.~Rhodes\orcid{0000-0002-4485-8549}\inst{\ref{aff67}}
\and G.~Riccio\inst{\ref{aff9}}
\and I.~Risso\orcid{0000-0003-2525-7761}\inst{\ref{aff32},\ref{aff33},\ref{aff25}}
\and F.~Rizzo\orcid{0000-0002-9407-585X}\inst{\ref{aff27}}
\and E.~Romelli\orcid{0000-0003-3069-9222}\inst{\ref{aff27}}
\and B.~Rusholme\orcid{0000-0001-7648-4142}\inst{\ref{aff66}}
\and R.~Saglia\orcid{0000-0003-0378-7032}\inst{\ref{aff63},\ref{aff62}}
\and Z.~Sakr\orcid{0000-0002-4823-3757}\inst{\ref{aff99},\ref{aff100},\ref{aff101}}
\and D.~Sapone\orcid{0000-0001-7089-4503}\inst{\ref{aff102}}
\and M.~Schirmer\orcid{0000-0003-2568-9994}\inst{\ref{aff72}}
\and P.~Schneider\orcid{0000-0001-8561-2679}\inst{\ref{aff79}}
\and T.~Schrabback\orcid{0000-0002-6987-7834}\inst{\ref{aff103}}
\and A.~Secroun\orcid{0000-0003-0505-3710}\inst{\ref{aff55}}
\and E.~Sihvola\orcid{0000-0003-1804-7715}\inst{\ref{aff104}}
\and P.~Simon\inst{\ref{aff79}}
\and C.~Sirignano\orcid{0000-0002-0995-7146}\inst{\ref{aff98},\ref{aff54}}
\and G.~Sirri\orcid{0000-0003-2626-2853}\inst{\ref{aff18}}
\and L.~Stanco\orcid{0000-0002-9706-5104}\inst{\ref{aff54}}
\and P.~Tallada-Cresp\'{i}\orcid{0000-0002-1336-8328}\inst{\ref{aff105},\ref{aff88}}
\and A.~N.~Taylor\inst{\ref{aff46}}
\and I.~Tereno\orcid{0000-0002-4537-6218}\inst{\ref{aff106},\ref{aff107}}
\and N.~Tessore\orcid{0000-0002-9696-7931}\inst{\ref{aff51}}
\and S.~Toft\orcid{0000-0003-3631-7176}\inst{\ref{aff108},\ref{aff109}}
\and R.~Toledo-Moreo\orcid{0000-0002-2997-4859}\inst{\ref{aff110},\ref{aff111}}
\and F.~Torradeflot\orcid{0000-0003-1160-1517}\inst{\ref{aff88},\ref{aff105}}
\and A.~Tsyganov\inst{\ref{aff112}}
\and I.~Tutusaus\orcid{0000-0002-3199-0399}\inst{\ref{aff59},\ref{aff58},\ref{aff100}}
\and J.~Valiviita\orcid{0000-0001-6225-3693}\inst{\ref{aff11},\ref{aff75}}
\and T.~Vassallo\orcid{0000-0001-6512-6358}\inst{\ref{aff27},\ref{aff61}}
\and Y.~Wang\orcid{0000-0002-4749-2984}\inst{\ref{aff66}}
\and J.~Weller\orcid{0000-0002-8282-2010}\inst{\ref{aff63},\ref{aff62}}
\and A.~Zacchei\orcid{0000-0003-0396-1192}\inst{\ref{aff27},\ref{aff26}}
\and G.~Zamorani\orcid{0000-0002-2318-301X}\inst{\ref{aff17}}
\and F.~M.~Zerbi\orcid{0000-0002-9996-973X}\inst{\ref{aff25}}
\and E.~Zucca\orcid{0000-0002-5845-8132}\inst{\ref{aff17}}
\and I.~T.~Andika\orcid{0000-0001-6102-9526}\inst{\ref{aff63}}
\and M.~Ballardini\orcid{0000-0003-4481-3559}\inst{\ref{aff113},\ref{aff114},\ref{aff17}}
\and C.~Benoist\inst{\ref{aff115}}
\and S.~Borgani\orcid{0000-0001-6151-6439}\inst{\ref{aff85},\ref{aff26},\ref{aff27},\ref{aff28},\ref{aff116}}
\and E.~Bozzo\orcid{0000-0002-8201-1525}\inst{\ref{aff52}}
\and C.~Burigana\orcid{0000-0002-3005-5796}\inst{\ref{aff117},\ref{aff57}}
\and R.~Cabanac\orcid{0000-0001-6679-2600}\inst{\ref{aff100}}
\and A.~Cappi\inst{\ref{aff115},\ref{aff17}}
\and F.~Caro\orcid{0009-0003-1053-0507}\inst{\ref{aff42}}
\and T.~Castro\orcid{0000-0002-6292-3228}\inst{\ref{aff118},\ref{aff27},\ref{aff26}}
\and J.~A.~Escartin~Vigo\inst{\ref{aff62}}
\and J.~Garc\'ia-Bellido\orcid{0000-0002-9370-8360}\inst{\ref{aff99}}
\and T.~Gasparetto\orcid{0000-0002-7913-4866}\inst{\ref{aff42}}
\and C.~M.~Gutierrez\orcid{0000-0001-7854-783X}\inst{\ref{aff45},\ref{aff119}}
\and A.~Hall\orcid{0000-0002-3139-8651}\inst{\ref{aff46}}
\and Y.~Kang\orcid{0009-0000-8588-7250}\inst{\ref{aff103}}
\and J.~Macias-Perez\orcid{0000-0002-5385-2763}\inst{\ref{aff120}}
\and R.~Maoli\orcid{0000-0002-6065-3025}\inst{\ref{aff121},\ref{aff42}}
\and M.~P\"ontinen\orcid{0000-0001-5442-2530}\inst{\ref{aff11}}
\and E.~Sarpa\orcid{0000-0002-1256-655X}\inst{\ref{aff27}}
\and V.~Scottez\orcid{0009-0008-3864-940X}\inst{\ref{aff122},\ref{aff123}}
\and M.~Sereno\orcid{0000-0003-0302-0325}\inst{\ref{aff17},\ref{aff18}}
\and K.~Tanidis\orcid{0000-0001-9843-5130}\inst{\ref{aff124}}
\and M.~Tenti\orcid{0000-0002-4254-5901}\inst{\ref{aff18}}
\and M.~Tucci\inst{\ref{aff52}}
\and M.~Wiesmann\orcid{0009-0000-8199-5860}\inst{\ref{aff64}}
\and J.~A.~Acevedo~Barroso\orcid{0000-0002-9654-1711}\inst{\ref{aff67}}
\and Y.~Akrami\orcid{0000-0002-2407-7956}\inst{\ref{aff99},\ref{aff125}}
\and G.~Angora\orcid{0000-0002-0316-6562}\inst{\ref{aff9},\ref{aff113}}
\and M.~Archidiacono\orcid{0000-0003-4952-9012}\inst{\ref{aff126},\ref{aff127}}
\and G.~Aric\`o\orcid{0000-0002-2802-2928}\inst{\ref{aff18}}
\and F.~Atrio-Barandela\orcid{0000-0002-2130-2513}\inst{\ref{aff128}}
\and M.~Baes\orcid{0000-0002-3930-2757}\inst{\ref{aff5}}
\and L.~Bazzanini\orcid{0000-0003-0727-0137}\inst{\ref{aff113},\ref{aff17}}
\and J.~R.~Bermejo-Climent\inst{\ref{aff129},\ref{aff130}}
\and D.~Bertacca\orcid{0000-0002-2490-7139}\inst{\ref{aff98},\ref{aff31},\ref{aff54}}
\and M.~Bethermin\orcid{0000-0002-3915-2015}\inst{\ref{aff131}}
\and F.~Beutler\orcid{0000-0003-0467-5438}\inst{\ref{aff46}}
\and A.~Blanchard\orcid{0000-0001-8555-9003}\inst{\ref{aff100}}
\and H.~B\"ohringer\orcid{0000-0001-8241-4204}\inst{\ref{aff62},\ref{aff61},\ref{aff132}}
\and F.~Bouch\`e\orcid{0000-0002-4663-1786}\inst{\ref{aff133},\ref{aff43}}
\and M.~L.~Brown\orcid{0000-0002-0370-8077}\inst{\ref{aff47}}
\and S.~Bruton\orcid{0000-0002-6503-5218}\inst{\ref{aff65}}
\and A.~Calabro\orcid{0000-0003-2536-1614}\inst{\ref{aff42}}
\and C.~S.~Carvalho\inst{\ref{aff134}}
\and F.~Cogato\orcid{0000-0003-4632-6113}\inst{\ref{aff17}}
\and P.~Corcho-Caballero\orcid{0000-0001-6327-7080}\inst{\ref{aff135}}
\and M.~Correnti\orcid{0000-0001-6464-3257}\inst{\ref{aff42},\ref{aff95}}
\and B.~Csizi\orcid{0000-0003-3227-6581}\inst{\ref{aff103}}
\and O.~Cucciati\orcid{0000-0002-9336-7551}\inst{\ref{aff17}}
\and T.~de~Boer\orcid{0000-0001-5486-2747}\inst{\ref{aff13}}
\and F.~De~Paolis\orcid{0000-0001-6460-7563}\inst{\ref{aff136},\ref{aff137},\ref{aff138}}
\and G.~Desprez\orcid{0000-0001-8325-1742}\inst{\ref{aff135}}
\and A.~D\'iaz-S\'anchez\orcid{0000-0003-0748-4768}\inst{\ref{aff139}}
\and S.~Di~Domizio\orcid{0000-0003-2863-5895}\inst{\ref{aff32},\ref{aff33}}
\and J.~M.~Diego\orcid{0000-0001-9065-3926}\inst{\ref{aff140}}
\and P.-A.~Duc\orcid{0000-0003-3343-6284}\inst{\ref{aff131}}
\and V.~Duret\orcid{0009-0009-0383-4960}\inst{\ref{aff55}}
\and M.~Y.~Elkhashab\orcid{0000-0001-9306-2603}\inst{\ref{aff27},\ref{aff28},\ref{aff85},\ref{aff26}}
\and Y.~Fang\orcid{0000-0002-0334-6950}\inst{\ref{aff63}}
\and A.~Farina\orcid{0009-0000-3420-929X}\inst{\ref{aff25},\ref{aff33}}
\and A.~Finoguenov\orcid{0000-0002-4606-5403}\inst{\ref{aff11}}
\and F.~Fontanot\orcid{0000-0003-4744-0188}\inst{\ref{aff27},\ref{aff26}}
\and A.~Franco\orcid{0000-0002-4761-366X}\inst{\ref{aff136},\ref{aff137},\ref{aff138}}
\and A.~Fumagalli\orcid{0009-0004-0300-2535}\inst{\ref{aff27}}
\and Y.~Fu\orcid{0000-0002-0759-0504}\inst{\ref{aff41},\ref{aff135}}
\and K.~Ganga\orcid{0000-0001-8159-8208}\inst{\ref{aff82}}
\and E.~Gaztanaga\orcid{0000-0001-9632-0815}\inst{\ref{aff59},\ref{aff58},\ref{aff7}}
\and Z.~Ghaffari\orcid{0000-0002-6467-8078}\inst{\ref{aff27},\ref{aff26}}
\and F.~Giacomini\orcid{0000-0002-3129-2814}\inst{\ref{aff18}}
\and F.~Gianotti\orcid{0000-0003-4666-119X}\inst{\ref{aff17}}
\and G.~Gozaliasl\orcid{0000-0002-0236-919X}\inst{\ref{aff141},\ref{aff11}}
\and A.~Gruppuso\orcid{0000-0001-9272-5292}\inst{\ref{aff17},\ref{aff18}}
\and M.~Guidi\orcid{0000-0001-9408-1101}\inst{\ref{aff30},\ref{aff17}}
\and H.~Hildebrandt\orcid{0000-0002-9814-3338}\inst{\ref{aff142}}
\and J.~Hjorth\orcid{0000-0002-4571-2306}\inst{\ref{aff143}}
\and M.~Jauzac\orcid{0000-0003-1974-8732}\inst{\ref{aff144},\ref{aff81},\ref{aff145},\ref{aff146},\ref{aff100}}
\and J.~J.~E.~Kajava\orcid{0000-0002-3010-8333}\inst{\ref{aff147},\ref{aff10},\ref{aff148}}
\and V.~Kansal\orcid{0000-0002-4008-6078}\inst{\ref{aff149},\ref{aff150}}
\and D.~Karagiannis\orcid{0000-0002-4927-0816}\inst{\ref{aff151},\ref{aff113},\ref{aff152}}
\and J.~Kim\orcid{0000-0003-2776-2761}\inst{\ref{aff60}}
\and C.~C.~Kirkpatrick\inst{\ref{aff104}}
\and A.~Kov\'acs\orcid{0000-0002-5825-579X}\inst{\ref{aff130},\ref{aff129}}
\and I.~Kova{\v{c}}i{\'{c}}\orcid{0000-0001-6751-3263}\inst{\ref{aff5}}
\and K.~Koyama\orcid{0000-0001-6727-6915}\inst{\ref{aff7}}
\and S.~Kruk\orcid{0000-0001-8010-8879}\inst{\ref{aff24}}
\and M.~C.~Lam\orcid{0000-0002-9347-2298}\inst{\ref{aff46}}
\and M.~Lattanzi\orcid{0000-0003-1059-2532}\inst{\ref{aff114}}
\and S.~Lee\orcid{0000-0002-8289-740X}\inst{\ref{aff153},\ref{aff67}}
\and M.~D.~Lepinzan\orcid{0000-0003-1287-9801}\inst{\ref{aff27}}
\and G.~Leroy\orcid{0009-0004-2523-4425}\inst{\ref{aff144},\ref{aff81}}
\and G.~F.~Lesci\orcid{0000-0002-4607-2830}\inst{\ref{aff16},\ref{aff117}}
\and J.~Lesgourgues\orcid{0000-0001-7627-353X}\inst{\ref{aff154}}
\and L.~Linke\orcid{0000-0002-2622-8113}\inst{\ref{aff103}}
\and G.~Maggio\orcid{0000-0003-4020-4836}\inst{\ref{aff27}}
\and A.~Manj\'on-Garc\'ia\orcid{0000-0002-7413-8825}\inst{\ref{aff139}}
\and F.~Mannucci\orcid{0000-0002-4803-2381}\inst{\ref{aff155}}
\and F.~R.~Marleau\orcid{0000-0002-1442-2947}\inst{\ref{aff103}}
\and C.~J.~A.~P.~Martins\orcid{0000-0002-4886-9261}\inst{\ref{aff156},\ref{aff35}}
\and M.~Miluzio\inst{\ref{aff24},\ref{aff157}}
\and G.~Morgante\inst{\ref{aff17}}
\and K.~Naidoo\orcid{0000-0002-9182-1802}\inst{\ref{aff7},\ref{aff72}}
\and A.~Navarro-Alsina\orcid{0000-0002-3173-2592}\inst{\ref{aff79}}
\and S.~Nesseris\orcid{0000-0002-0567-0324}\inst{\ref{aff99}}
\and L.~Nicastro\orcid{0000-0001-8534-6788}\inst{\ref{aff17}}
\and F.~Oppizzi\orcid{0000-0003-3904-8370}\inst{\ref{aff33},\ref{aff54},\ref{aff98}}
\and F.~Pace\orcid{0000-0001-8039-0480}\inst{\ref{aff38},\ref{aff39},\ref{aff40},\ref{aff107}}
\and D.~Paoletti\orcid{0000-0003-4761-6147}\inst{\ref{aff17},\ref{aff57}}
\and K.~Paterson\orcid{0000-0001-8340-3486}\inst{\ref{aff72}}
\and L.~Patrizii\inst{\ref{aff18}}
\and C.~Pattison\orcid{0000-0003-3272-2617}\inst{\ref{aff7}}
\and A.~Pisani\orcid{0000-0002-6146-4437}\inst{\ref{aff55}}
\and L.~Pizzuti\orcid{0000-0001-5654-7580}\inst{\ref{aff158}}
\and D.~Potter\orcid{0000-0002-0757-5195}\inst{\ref{aff159}}
\and G.~W.~Pratt\inst{\ref{aff94}}
\and S.~Quai\orcid{0000-0002-0449-8163}\inst{\ref{aff16},\ref{aff17}}
\and M.~Radovich\orcid{0000-0002-3585-866X}\inst{\ref{aff31}}
\and F.~Ricci\orcid{0000-0001-5742-5980}\inst{\ref{aff160},\ref{aff42}}
\and W.~Roster\orcid{0000-0002-9149-6528}\inst{\ref{aff62}}
\and M.~Sahl\'en\orcid{0000-0003-0973-4804}\inst{\ref{aff161}}
\and D.~B.~Sanders\orcid{0000-0002-1233-9998}\inst{\ref{aff13}}
\and C.~Scarlata\orcid{0000-0002-9136-8876}\inst{\ref{aff162}}
\and J.~Schaye\orcid{0000-0002-0668-5560}\inst{\ref{aff41}}
\and A.~Schneider\orcid{0000-0001-7055-8104}\inst{\ref{aff159}}
\and D.~Sciotti\orcid{0009-0008-4519-2620}\inst{\ref{aff42},\ref{aff80}}
\and E.~Sellentin\orcid{0009-0002-2655-3458}\inst{\ref{aff163},\ref{aff41}}
\and L.~C.~Smith\orcid{0000-0002-3259-2771}\inst{\ref{aff164}}
\and J.~G.~Sorce\orcid{0000-0002-2307-2432}\inst{\ref{aff165},\ref{aff53}}
\and I.~Szapudi\orcid{0000-0003-2274-0301}\inst{\ref{aff13}}
\and M.~Talia\orcid{0000-0003-4352-2063}\inst{\ref{aff16},\ref{aff17}}
\and F.~Tarsitano\orcid{0000-0002-5919-0238}\inst{\ref{aff166},\ref{aff167}}
\and G.~Testera\orcid{0000-0003-2970-766X}\inst{\ref{aff33}}
\and R.~Teyssier\orcid{0000-0001-7689-0933}\inst{\ref{aff168}}
\and S.~Tosi\orcid{0000-0002-7275-9193}\inst{\ref{aff32},\ref{aff25},\ref{aff33}}
\and A.~Troja\orcid{0000-0003-0239-4595}\inst{\ref{aff27}}
\and C.~Uhlemann\orcid{0000-0001-7831-1579}\inst{\ref{aff169},\ref{aff8}}
\and C.~Valieri\inst{\ref{aff18}}
\and A.~Venhola\orcid{0000-0001-6071-4564}\inst{\ref{aff170}}
\and D.~Vergani\orcid{0000-0003-0898-2216}\inst{\ref{aff17}}
\and G.~Vernardos\orcid{0000-0001-8554-7248}\inst{\ref{aff171},\ref{aff172}}
\and E.~Vilenius\orcid{0000-0002-6184-7681}\inst{\ref{aff51}}
\and S.~Vinciguerra\orcid{0009-0005-4018-3184}\inst{\ref{aff14}}
\and M.~von~Wietersheim-Kramsta\orcid{0000-0003-4986-5091}\inst{\ref{aff81},\ref{aff144}}
\and N.~A.~Walton\orcid{0000-0003-3983-8778}\inst{\ref{aff164}}
\and A.~H.~Wright\orcid{0000-0001-7363-7932}\inst{\ref{aff142}}
\and H.~W.~Yeung\orcid{0000-0002-4993-9014}\inst{\ref{aff46}}
\and D.~Scott\orcid{0000-0002-6878-9840}\inst{\ref{aff173}}}
										   
\institute{Department of Astronomy, School of Physics and Astronomy, Shanghai Jiao Tong University, Shanghai 200240, China\label{aff1}
\and
State Key Laboratory of Dark Matter Physics, School of Physics and Astronomy, Shanghai Jiao Tong University, Shanghai 200240, China\label{aff2}
\and
Shanghai Key Laboratory for Particle Physics and Cosmology, Shanghai 200240, China\label{aff3}
\and
STAR Institute, University of Li{\`e}ge, Quartier Agora, All\'ee du six Ao\^ut 19c, 4000 Li\`ege, Belgium\label{aff4}
\and
Universiteit Gent, Department of Physics and Astronomy, Proeftuinstraat 86 N3, 9000 Ghent, Belgium
\label{aff5}
\and
University of Applied Sciences and Arts of Northwestern Switzerland, School of Computer Science, 5210 Windisch, Switzerland\label{aff6}
\and
Institute of Cosmology and Gravitation, University of Portsmouth, Portsmouth PO1 3FX, UK\label{aff7}
\and
School of Mathematics, Statistics and Physics, Newcastle University, Herschel Building, Newcastle-upon-Tyne, NE1 7RU, UK\label{aff8}
\and
INAF-Osservatorio Astronomico di Capodimonte, Via Moiariello 16, 80131 Napoli, Italy\label{aff9}
\and
Finnish Centre for Astronomy with ESO (FINCA), Quantum, Vesilinnantie 5, University of Turku, 20014 Turku, Finland\label{aff10}
\and
Department of Physics, P.O. Box 64, University of Helsinki, 00014 Helsinki, Finland\label{aff11}
\and
INAF-IASF Milano, Via Alfonso Corti 12, 20133 Milano, Italy\label{aff12}
\and
Institute for Astronomy, University of Hawaii, 2680 Woodlawn Drive, Honolulu, HI 96822, USA\label{aff13}
\and
Aix-Marseille Universit\'e, CNRS, CNES, LAM, Marseille, France\label{aff14}
\and
Institut d'Astrophysique de Paris, UMR 7095, CNRS, and Sorbonne Universit\'e, 98 bis boulevard Arago, 75014 Paris, France\label{aff15}
\and
Dipartimento di Fisica e Astronomia "Augusto Righi" - Alma Mater Studiorum Universit\`a di Bologna, via Piero Gobetti 93/2, 40129 Bologna, Italy\label{aff16}
\and
INAF-Osservatorio di Astrofisica e Scienza dello Spazio di Bologna, Via Piero Gobetti 93/3, 40129 Bologna, Italy\label{aff17}
\and
INFN-Sezione di Bologna, Viale Berti Pichat 6/2, 40127 Bologna, Italy\label{aff18}
\and
School of Physical Sciences, The Open University, Milton Keynes, MK7 6AA, UK\label{aff19}
\and
Institut de Ci\`{e}ncies del Cosmos (ICCUB), Universitat de Barcelona (IEEC-UB), Mart\'{i} i Franqu\`{e}s 1, 08028 Barcelona, Spain\label{aff20}
\and
Instituci\'o Catalana de Recerca i Estudis Avan\c{c}ats (ICREA), Passeig de Llu\'{\i}s Companys 23, 08010 Barcelona, Spain\label{aff21}
\and
Institut de Ciencies de l'Espai (IEEC-CSIC), Campus UAB, Carrer de Can Magrans, s/n Cerdanyola del Vall\'es, 08193 Barcelona, Spain\label{aff22}
\and
Max-Planck-Institut f\"ur Astrophysik, Karl-Schwarzschild-Str.~1, 85748 Garching, Germany\label{aff23}
\and
ESAC/ESA, Camino Bajo del Castillo, s/n., Urb. Villafranca del Castillo, 28692 Villanueva de la Ca\~nada, Madrid, Spain\label{aff24}
\and
INAF-Osservatorio Astronomico di Brera, Via Brera 28, 20122 Milano, Italy\label{aff25}
\and
IFPU, Institute for Fundamental Physics of the Universe, via Beirut 2, 34151 Trieste, Italy\label{aff26}
\and
INAF-Osservatorio Astronomico di Trieste, Via G. B. Tiepolo 11, 34143 Trieste, Italy\label{aff27}
\and
INFN, Sezione di Trieste, Via Valerio 2, 34127 Trieste TS, Italy\label{aff28}
\and
SISSA, International School for Advanced Studies, Via Bonomea 265, 34136 Trieste TS, Italy\label{aff29}
\and
Dipartimento di Fisica e Astronomia, Universit\`a di Bologna, Via Gobetti 93/2, 40129 Bologna, Italy\label{aff30}
\and
INAF-Osservatorio Astronomico di Padova, Via dell'Osservatorio 5, 35122 Padova, Italy\label{aff31}
\and
Dipartimento di Fisica, Universit\`a di Genova, Via Dodecaneso 33, 16146, Genova, Italy\label{aff32}
\and
INFN-Sezione di Genova, Via Dodecaneso 33, 16146, Genova, Italy\label{aff33}
\and
Department of Physics "E. Pancini", University Federico II, Via Cinthia 6, 80126, Napoli, Italy\label{aff34}
\and
Instituto de Astrof\'isica e Ci\^encias do Espa\c{c}o, Universidade do Porto, CAUP, Rua das Estrelas, PT4150-762 Porto, Portugal\label{aff35}
\and
Faculdade de Ci\^encias da Universidade do Porto, Rua do Campo de Alegre, 4150-007 Porto, Portugal\label{aff36}
\and
European Southern Observatory, Karl-Schwarzschild-Str.~2, 85748 Garching, Germany\label{aff37}
\and
Dipartimento di Fisica, Universit\`a degli Studi di Torino, Via P. Giuria 1, 10125 Torino, Italy\label{aff38}
\and
INFN-Sezione di Torino, Via P. Giuria 1, 10125 Torino, Italy\label{aff39}
\and
INAF-Osservatorio Astrofisico di Torino, Via Osservatorio 20, 10025 Pino Torinese (TO), Italy\label{aff40}
\and
Leiden Observatory, Leiden University, Einsteinweg 55, 2333 CC Leiden, The Netherlands\label{aff41}
\and
INAF-Osservatorio Astronomico di Roma, Via Frascati 33, 00078 Monteporzio Catone, Italy\label{aff42}
\and
INFN -- Sezione di Napoli, Via Cinthia 6, 80126, Napoli, Italy\label{aff43}
\and
Dipartimento di Fisica e Astronomia "Augusto Righi" - Alma Mater Studiorum Universit\`a di Bologna, Viale Berti Pichat 6/2, 40127 Bologna, Italy\label{aff44}
\and
Instituto de Astrof\'{\i}sica de Canarias, E-38205 La Laguna, Tenerife, Spain\label{aff45}
\and
Institute for Astronomy, University of Edinburgh, Royal Observatory, Blackford Hill, Edinburgh EH9 3HJ, UK\label{aff46}
\and
Jodrell Bank Centre for Astrophysics, Department of Physics and Astronomy, University of Manchester, Oxford Road, Manchester M13 9PL, UK\label{aff47}
\and
European Space Agency/ESRIN, Largo Galileo Galilei 1, 00044 Frascati, Roma, Italy\label{aff48}
\and
Universit\'e Claude Bernard Lyon 1, CNRS/IN2P3, IP2I Lyon, UMR 5822, Villeurbanne, F-69100, France\label{aff49}
\and
UCB Lyon 1, CNRS/IN2P3, IUF, IP2I Lyon, 4 rue Enrico Fermi, 69622 Villeurbanne, France\label{aff50}
\and
Mullard Space Science Laboratory, University College London, Holmbury St Mary, Dorking, Surrey RH5 6NT, UK\label{aff51}
\and
Department of Astronomy, University of Geneva, ch. d'Ecogia 16, 1290 Versoix, Switzerland\label{aff52}
\and
Universit\'e Paris-Saclay, CNRS, Institut d'astrophysique spatiale, 91405, Orsay, France\label{aff53}
\and
INFN-Padova, Via Marzolo 8, 35131 Padova, Italy\label{aff54}
\and
Aix-Marseille Universit\'e, CNRS/IN2P3, CPPM, Marseille, France\label{aff55}
\and
INAF-Istituto di Astrofisica e Planetologia Spaziali, via del Fosso del Cavaliere, 100, 00100 Roma, Italy\label{aff56}
\and
INFN-Bologna, Via Irnerio 46, 40126 Bologna, Italy\label{aff57}
\and
Institut d'Estudis Espacials de Catalunya (IEEC),  Edifici RDIT, Campus UPC, 08860 Castelldefels, Barcelona, Spain\label{aff58}
\and
Institute of Space Sciences (ICE, CSIC), Campus UAB, Carrer de Can Magrans, s/n, 08193 Barcelona, Spain\label{aff59}
\and
Department of Physics, University of Oxford, Keble Road, Oxford OX1 3RH, UK\label{aff60}
\and
University Observatory, LMU Faculty of Physics, Scheinerstr.~1, 81679 Munich, Germany\label{aff61}
\and
Max Planck Institute for Extraterrestrial Physics, Giessenbachstr. 1, 85748 Garching, Germany\label{aff62}
\and
Universit\"ats-Sternwarte M\"unchen, Fakult\"at f\"ur Physik, Ludwig-Maximilians-Universit\"at M\"unchen, Scheinerstr.~1, 81679 M\"unchen, Germany\label{aff63}
\and
Institute of Theoretical Astrophysics, University of Oslo, P.O. Box 1029 Blindern, 0315 Oslo, Norway\label{aff64}
\and
California Institute of Technology, 1200 E California Blvd, Pasadena, CA 91125, USA\label{aff65}
\and
Caltech/IPAC, 1200 E. California Blvd., Pasadena, CA 91125, USA\label{aff66}
\and
Jet Propulsion Laboratory, California Institute of Technology, 4800 Oak Grove Drive, Pasadena, CA, 91109, USA\label{aff67}
\and
Department of Physics, Lancaster University, Lancaster, LA1 4YB, UK\label{aff68}
\and
Felix Hormuth Engineering, Goethestr. 17, 69181 Leimen, Germany\label{aff69}
\and
Technical University of Denmark, Elektrovej 327, 2800 Kgs. Lyngby, Denmark\label{aff70}
\and
Cosmic Dawn Center (DAWN), Denmark\label{aff71}
\and
Max-Planck-Institut f\"ur Astronomie, K\"onigstuhl 17, 69117 Heidelberg, Germany\label{aff72}
\and
NASA Goddard Space Flight Center, Greenbelt, MD 20771, USA\label{aff73}
\and
Universit\'e de Gen\`eve, D\'epartement de Physique Th\'eorique and Centre for Astroparticle Physics, 24 quai Ernest-Ansermet, CH-1211 Gen\`eve 4, Switzerland\label{aff74}
\and
Helsinki Institute of Physics, Gustaf H{\"a}llstr{\"o}min katu 2, University of Helsinki, 00014 Helsinki, Finland\label{aff75}
\and
Laboratoire d'etude de l'Univers et des phenomenes eXtremes, Observatoire de Paris, Universit\'e PSL, Sorbonne Universit\'e, CNRS, 92190 Meudon, France\label{aff76}
\and
SKAO, Jodrell Bank, Lower Withington, Macclesfield SK11 9FT, UK\label{aff77}
\and
Centre de Calcul de l'IN2P3/CNRS, 21 avenue Pierre de Coubertin 69627 Villeurbanne Cedex, France\label{aff78}
\and
Universit\"at Bonn, Argelander-Institut f\"ur Astronomie, Auf dem H\"ugel 71, 53121 Bonn, Germany\label{aff79}
\and
INFN-Sezione di Roma, Piazzale Aldo Moro, 2 - c/o Dipartimento di Fisica, Edificio G. Marconi, 00185 Roma, Italy\label{aff80}
\and
Department of Physics, Institute for Computational Cosmology, Durham University, South Road, Durham, DH1 3LE, UK\label{aff81}
\and
Universit\'e Paris Cit\'e, CNRS, Astroparticule et Cosmologie, 75013 Paris, France\label{aff82}
\and
CNRS-UCB International Research Laboratory, Centre Pierre Bin\'etruy, IRL2007, CPB-IN2P3, Berkeley, USA\label{aff83}
\and
Institute of Physics, Laboratory of Astrophysics, Ecole Polytechnique F\'ed\'erale de Lausanne (EPFL), Observatoire de Sauverny, 1290 Versoix, Switzerland\label{aff84}
\and
Dipartimento di Fisica - Sezione di Astronomia, Universit\`a di Trieste, Via Tiepolo 11, 34131 Trieste, Italy\label{aff85}
\and
Telespazio UK S.L. for European Space Agency (ESA), Camino bajo del Castillo, s/n, Urbanizacion Villafranca del Castillo, Villanueva de la Ca\~nada, 28692 Madrid, Spain\label{aff86}
\and
Institut de F\'{i}sica d'Altes Energies (IFAE), The Barcelona Institute of Science and Technology, Campus UAB, 08193 Bellaterra (Barcelona), Spain\label{aff87}
\and
Port d'Informaci\'{o} Cient\'{i}fica, Campus UAB, C. Albareda s/n, 08193 Bellaterra (Barcelona), Spain\label{aff88}
\and
School of Mathematics and Physics, University of Surrey, Guildford, Surrey, GU2 7XH, UK\label{aff89}
\and
European Space Agency/ESTEC, Keplerlaan 1, 2201 AZ Noordwijk, The Netherlands\label{aff90}
\and
Waterloo Centre for Astrophysics, University of Waterloo, Waterloo, Ontario N2L 3G1, Canada\label{aff91}
\and
Department of Physics and Astronomy, University of Waterloo, Waterloo, Ontario N2L 3G1, Canada\label{aff92}
\and
Perimeter Institute for Theoretical Physics, Waterloo, Ontario N2L 2Y5, Canada\label{aff93}
\and
Universit\'e Paris-Saclay, Universit\'e Paris Cit\'e, CEA, CNRS, AIM, 91191, Gif-sur-Yvette, France\label{aff94}
\and
Space Science Data Center, Italian Space Agency, via del Politecnico snc, 00133 Roma, Italy\label{aff95}
\and
Centre National d'Etudes Spatiales -- Centre spatial de Toulouse, 18 avenue Edouard Belin, 31401 Toulouse Cedex 9, France\label{aff96}
\and
Institute of Space Science, Str. Atomistilor, nr. 409 M\u{a}gurele, Ilfov, 077125, Romania\label{aff97}
\and
Dipartimento di Fisica e Astronomia "G. Galilei", Universit\`a di Padova, Via Marzolo 8, 35131 Padova, Italy\label{aff98}
\and
Instituto de F\'isica Te\'orica UAM-CSIC, Campus de Cantoblanco, 28049 Madrid, Spain\label{aff99}
\and
Institut de Recherche en Astrophysique et Plan\'etologie (IRAP), Universit\'e de Toulouse, CNRS, UPS, CNES, 14 Av. Edouard Belin, 31400 Toulouse, France\label{aff100}
\and
Universit\'e St Joseph; Faculty of Sciences, Beirut, Lebanon\label{aff101}
\and
Departamento de F\'isica, FCFM, Universidad de Chile, Blanco Encalada 2008, Santiago, Chile\label{aff102}
\and
Universit\"at Innsbruck, Institut f\"ur Astro- und Teilchenphysik, Technikerstr. 25/8, 6020 Innsbruck, Austria\label{aff103}
\and
Department of Physics and Helsinki Institute of Physics, Gustaf H\"allstr\"omin katu 2, University of Helsinki, 00014 Helsinki, Finland\label{aff104}
\and
Centro de Investigaciones Energ\'eticas, Medioambientales y Tecnol\'ogicas (CIEMAT), Avenida Complutense 40, 28040 Madrid, Spain\label{aff105}
\and
Departamento de F\'isica, Faculdade de Ci\^encias, Universidade de Lisboa, Edif\'icio C8, Campo Grande, PT1749-016 Lisboa, Portugal\label{aff106}
\and
Instituto de Astrof\'isica e Ci\^encias do Espa\c{c}o, Faculdade de Ci\^encias, Universidade de Lisboa, Campo Grande, 1749-016 Lisboa, Portugal\label{aff107}
\and
Cosmic Dawn Center (DAWN)\label{aff108}
\and
Niels Bohr Institute, University of Copenhagen, Jagtvej 128, 2200 Copenhagen, Denmark\label{aff109}
\and
Universidad Polit\'ecnica de Cartagena, Departamento de Electr\'onica y Tecnolog\'ia de Computadoras,  Plaza del Hospital 1, 30202 Cartagena, Spain\label{aff110}
\and
European University of Technology EUt+, European Union\label{aff111}
\and
Centre for Information Technology, University of Groningen, P.O. Box 11044, 9700 CA Groningen, The Netherlands\label{aff112}
\and
Dipartimento di Fisica e Scienze della Terra, Universit\`a degli Studi di Ferrara, Via Giuseppe Saragat 1, 44122 Ferrara, Italy\label{aff113}
\and
Istituto Nazionale di Fisica Nucleare, Sezione di Ferrara, Via Giuseppe Saragat 1, 44122 Ferrara, Italy\label{aff114}
\and
Universit\'e C\^{o}te d'Azur, Observatoire de la C\^{o}te d'Azur, CNRS, Laboratoire Lagrange, Bd de l'Observatoire, CS 34229, 06304 Nice cedex 4, France\label{aff115}
\and
ICSC - Centro Nazionale di Ricerca in High Performance Computing, Big Data e Quantum Computing, Via Magnanelli 2, Bologna, Italy\label{aff116}
\and
INAF, Istituto di Radioastronomia, Via Piero Gobetti 101, 40129 Bologna, Italy\label{aff117}
\and
Department of Mathematical Physics, Institute of Physics, University of S\~ao Paulo, R. do Mat\~ao 1371, 05508-090, S\~ao Paulo, SP, Brazil\label{aff118}
\and
Universidad de La Laguna, Dpto. Astrof\'\i sica, E-38206 La Laguna, Tenerife, Spain\label{aff119}
\and
Univ. Grenoble Alpes, CNRS, Grenoble INP, LPSC-IN2P3, 53, Avenue des Martyrs, 38000 Grenoble, France\label{aff120}
\and
Dipartimento di Fisica, Sapienza Universit\`a di Roma, Piazzale Aldo Moro 2, 00185 Roma, Italy\label{aff121}
\and
Institut d'Astrophysique de Paris, 98bis Boulevard Arago, 75014, Paris, France\label{aff122}
\and
ICL, Junia, Universit\'e Catholique de Lille, LITL, 59000 Lille, France\label{aff123}
\and
Center for Astrophysics and Cosmology, University of Nova Gorica, Nova Gorica, Slovenia\label{aff124}
\and
CERCA/ISO, Department of Physics, Case Western Reserve University, 10900 Euclid Avenue, Cleveland, OH 44106, USA\label{aff125}
\and
Dipartimento di Fisica "Aldo Pontremoli", Universit\`a degli Studi di Milano, Via Celoria 16, 20133 Milano, Italy\label{aff126}
\and
INFN-Sezione di Milano, Via Celoria 16, 20133 Milano, Italy\label{aff127}
\and
Departamento de F{\'\i}sica Fundamental. Universidad de Salamanca. Plaza de la Merced s/n. 37008 Salamanca, Spain\label{aff128}
\and
Konkoly Observatory, HUN-REN CSFK, MTA Centre of Excellence, Budapest, Konkoly Thege Mikl\'os {\'u}t 15-17. H-1121, Hungary\label{aff129}
\and
MTA-CSFK Lend\"ulet Large-Scale Structure Research Group, Konkoly-Thege Mikl\'os \'ut 15-17, H-1121 Budapest, Hungary\label{aff130}
\and
Universit\'e de Strasbourg, CNRS, Observatoire astronomique de Strasbourg, UMR 7550, 67000 Strasbourg, France\label{aff131}
\and
Max-Planck-Institut f\"ur Physik, Boltzmannstr. 8, 85748 Garching, Germany\label{aff132}
\and
Scuola Superiore Meridionale, Via Mezzocannone 4, 80138, Napoli, Italy\label{aff133}
\and
Instituto de Astrof\'isica e Ci\^encias do Espa\c{c}o, Faculdade de Ci\^encias, Universidade de Lisboa, Tapada da Ajuda, 1349-018 Lisboa, Portugal\label{aff134}
\and
Kapteyn Astronomical Institute, University of Groningen, PO Box 800, 9700 AV Groningen, The Netherlands\label{aff135}
\and
Department of Mathematics and Physics E. De Giorgi, University of Salento, Via per Arnesano, CP-I93, 73100, Lecce, Italy\label{aff136}
\and
INFN, Sezione di Lecce, Via per Arnesano, CP-193, 73100, Lecce, Italy\label{aff137}
\and
INAF-Sezione di Lecce, c/o Dipartimento Matematica e Fisica, Via per Arnesano, 73100, Lecce, Italy\label{aff138}
\and
Departamento F\'isica Aplicada, Universidad Polit\'ecnica de Cartagena, Campus Muralla del Mar, 30202 Cartagena, Murcia, Spain\label{aff139}
\and
Instituto de F\'isica de Cantabria, Edificio Juan Jord\'a, Avenida de los Castros, 39005 Santander, Spain\label{aff140}
\and
Department of Computer Science, Aalto University, PO Box 15400, Espoo, FI-00 076, Finland\label{aff141}
\and
Ruhr University Bochum, Faculty of Physics and Astronomy, Astronomical Institute (AIRUB), German Centre for Cosmological Lensing (GCCL), 44780 Bochum, Germany\label{aff142}
\and
DARK, Niels Bohr Institute, University of Copenhagen, Jagtvej 155, 2200 Copenhagen, Denmark\label{aff143}
\and
Department of Physics, Centre for Extragalactic Astronomy, Durham University, South Road, Durham, DH1 3LE, UK\label{aff144}
\and
Astrophysics Research Centre, University of KwaZulu-Natal, Westville Campus, Durban 4041, South Africa\label{aff145}
\and
School of Mathematics, Statistics \& Computer Science, University of KwaZulu-Natal, Westville Campus, Durban 4041, South Africa\label{aff146}
\and
Department of Physics and Astronomy, Vesilinnantie 5, University of Turku, 20014 Turku, Finland\label{aff147}
\and
Serco for European Space Agency (ESA), Camino bajo del Castillo, s/n, Urbanizacion Villafranca del Castillo, Villanueva de la Ca\~nada, 28692 Madrid, Spain\label{aff148}
\and
ARC Centre of Excellence for Dark Matter Particle Physics, Melbourne, Australia\label{aff149}
\and
Centre for Astrophysics \& Supercomputing, Swinburne University of Technology,  Hawthorn, Victoria 3122, Australia\label{aff150}
\and
Van Swinderen Institute, University of Groningen, Nijenborgh 3, 9747 AG Groningen, The Netherlands\label{aff151}
\and
Department of Physics and Astronomy, University of the Western Cape, Bellville, Cape Town, 7535, South Africa\label{aff152}
\and
Ohio University, Physics \& Astronomy Department,1 Ohio University, Athens, OH 45701, USA\label{aff153}
\and
Institute for Theoretical Particle Physics and Cosmology (TTK), RWTH Aachen University, 52056 Aachen, Germany\label{aff154}
\and
INAF-Osservatorio Astrofisico di Arcetri, Largo E. Fermi 5, 50125, Firenze, Italy\label{aff155}
\and
Centro de Astrof\'{\i}sica da Universidade do Porto, Rua das Estrelas, 4150-762 Porto, Portugal\label{aff156}
\and
HE Space for European Space Agency (ESA), Camino bajo del Castillo, s/n, Urbanizacion Villafranca del Castillo, Villanueva de la Ca\~nada, 28692 Madrid, Spain\label{aff157}
\and
Dipartimento di Fisica ``G. Occhialini", Universit\`a degli Studi di Milano Bicocca, Piazza della Scienza 3, 20126 Milano, Italy\label{aff158}
\and
Department of Astrophysics, University of Zurich, Winterthurerstrasse 190, 8057 Zurich, Switzerland\label{aff159}
\and
Department of Mathematics and Physics, Roma Tre University, Via della Vasca Navale 84, 00146 Rome, Italy\label{aff160}
\and
Theoretical astrophysics, Department of Physics and Astronomy, Uppsala University, Box 524, 751 20 Uppsala, Sweden\label{aff161}
\and
Minnesota Institute for Astrophysics, University of Minnesota, 116 Church St SE, Minneapolis, MN 55455, USA\label{aff162}
\and
Mathematical Institute, University of Leiden, Einsteinweg 55, 2333 CA Leiden, The Netherlands\label{aff163}
\and
Institute of Astronomy, University of Cambridge, Madingley Road, Cambridge CB3 0HA, UK\label{aff164}
\and
Univ. Lille, CNRS, Centrale Lille, UMR 9189 CRIStAL, 59000 Lille, France\label{aff165}
\and
Kobayashi-Maskawa Institute for the Origin of Particles and the Universe, Nagoya University, Chikusa-ku, Nagoya, 464-8602, Japan\label{aff166}
\and
Institute for Particle Physics and Astrophysics, Dept. of Physics, ETH Zurich, Wolfgang-Pauli-Strasse 27, 8093 Zurich, Switzerland\label{aff167}
\and
Department of Astrophysical Sciences, Peyton Hall, Princeton University, Princeton, NJ 08544, USA\label{aff168}
\and
Fakult\"at f\"ur Physik, Universit\"at Bielefeld, Postfach 100131, 33501 Bielefeld, Germany\label{aff169}
\and
Space physics and astronomy research unit, University of Oulu, Pentti Kaiteran katu 1, FI-90014 Oulu, Finland\label{aff170}
\and
Department of Physics and Astronomy, Lehman College of the CUNY, Bronx, NY 10468, USA\label{aff171}
\and
American Museum of Natural History, Department of Astrophysics, New York, NY 10024, USA\label{aff172}
\and
Department of Physics and Astronomy, University of British Columbia, Vancouver, BC V6T 1Z1, Canada\label{aff173}}

%
\abstract{
Mass is among the most fundamental properties describing a galaxy.
The relation between the luminosity and the total mass of a galaxy contains information on the efficiency of the star formation process and on the distribution of dark matter.
We used strong gravitational lensing data from the Euclid Quick Data Release 1 (Q1) to constrain the total mass enclosed within an aperture of $\fivekpc$, $M_5$, of elliptical galaxies.
Our analysis is based on a sample of $\Ngold$ lenses, which we built by ensuring that their selection function is reproducible.
This enabled us to correct our inference for selection effects.
The average $\logten{(M_5\,h\,M_\odot^{-1})}$ for a galaxy with an $r$-band luminosity of $10^{10.7}L_{\odot}\,h^{-2}$ at $z=0.5$ is $11.30\pm0.02$. The mass $M_5$ scales with luminosity to the power $\beta_\mathrm{L}=0.71\pm0.07$, and decreases with increasing redshift at fixed luminosity. 
Assuming a scenario in which galaxies grow by mergers and stellar populations evolve passively, these correlations imply that $\mfive$ increases with time for individual galaxies.
These measurements provide a useful benchmark for testing models of galaxy structure and evolution that make predictions on the distribution of total mass on the scales of $\fivekpc$.
    }
%
%
%
%
   \titlerunning{The mass-to-light ratio of elliptical galaxies}
   \authorrunning{Euclid Collaboration: A. Sonnenfeld et al. }
   
   \maketitle
   \nolinenumbers
%
%
%
%
   
\section{\label{sc:Intro}Introduction}

Elliptical galaxies are some of the best-studied objects in the Universe. 
The properties of their luminous component are known in great detail. However, key aspects of their inner mass structure, such as the distribution of dark matter, are still uncertain. 
Mass is the quantity that is most robustly predicted in simulations of galaxies. Therefore, observations that directly trace mass offer a better chance for making accurate comparisons between theoretical models and observations. Models with different recipes for processes dominated by baryons, such as star formation or the ejection of gas from active galactic nuclei, make different predictions on the inner mass distribution \citep[e.g.][]{Duf++10,Pei++17}, and so do simulations with different models of dark matter \citep{Des++26}. Therefore, precise measurements of mass can help us to build more realistic models of galaxies.

Strong gravitational lensing offers the most direct means of obtaining measurements of mass in galaxies. 
A strong lens can provide a direct estimate of the total projected mass enclosed within the multiple images of the lensed background source, with a precision on the order of $1\%$.
Hundreds of such measurements have been obtained so far \citep{Aug++09,Bro++12,Son++13a,Shu++15,Old++17}. However, there are two aspects of strong lensing observations that complicate their interpretation. The first one is the fact that the Einstein radius, which is the main observable in strong lensing, is not purely an intrinsic property of the foreground lens galaxy, but also depends on the redshift of the background source.  Consequently, comparing theoretical models to strong lensing data is not as simple as doing this with more traditional observations, such as the stellar mass or the half-light radius. 
To overcome this obstacle, it is helpful to have a model for the density profile of the lens galaxies.
The most commonly adopted model of this type is a power law, in which the density is assumed to scale as $\rho(r)\propto r^{-\gamma}$.
The density slope $\gamma$ of early-type galaxies has been the focus of several strong lensing-based observational studies \citep{Koo++06,Aug++10,Bol++12,Son++13b,Tan++24,Sah++24}, and also of kinematics-only analyses \citep{Tor++14,Cap++15}.
Curiously, very little attention has been paid to the second parameter of the power-law model, the normalisation of the density\footnote{Historically, information on the overall mass has been conveyed in terms of the dark matter fraction within a given aperture \citep{Aug++10,Nap++10,Sha++21}.}. This will be the main focus of our work.

The second obstacle in the interpretation of strong lensing data is the fact that strong lenses are a biased subset of the general population of galaxies. More massive and more compact galaxies have a larger cross-section for creating strong lensing events, and are therefore overrepresented in strong lens samples \citep{Son++23}. 
Therefore, in order to generalise measurements obtained on strong lenses to the full population of galaxies, selection effects must be taken into account.

The \Euclid satellite \citep{EuclidSkyOverview} is revolutionising the field of strong lensing. 
Hundreds of new galaxy-scale strong lenses were discovered in the first  Euclid Quick Data Release \citep[Q1;][]{Q1cite,SLDEA,SLDEB,SLDEC,SLDED,SLDEE}. The Q1 data cover an area of $63$~deg$^2$ \citep{Q1-TP001}, meaning that more than $10^5$ lenses are expected to be discovered in the entire Euclid Wide Survey (which will ultimately cover an area of $14\,000$~deg$^2$).
In this work, we use the Q1 \Euclid strong lensing data to carry out a statistical study of the mass distribution in elliptical galaxies, including a detailed treatment of selection effects.
The Q1 strong lens sample is sufficiently large to produce competitive statistical measurements, while sufficiently small to allow for human-led lens modelling, which at this stage is a necessary step in our analysis. The results obtained in this work will help us guide in the analysis of the much larger sample of lenses from the Data Release 1.

The main quantity that we seek to constrain is the total projected mass enclosed within a cylinder of fixed physical radius. We set this radius to $\fivekpc$, as this is close to the average Einstein radius of the lenses in the \Euclid sample. 
We want to measure how the total fixed-aperture mass varies with luminosity, half-light radius, and redshift, across the population of elliptical galaxies. 
We choose to work with luminosity, rather than stellar mass, because it can be determined much more robustly. 
Our main result is therefore a measurement of the mass-to-light ratio integrated within an aperture, $\fivekpc$, where baryons and dark matter contribute with comparable amounts to the total mass of a typical massive elliptical galaxy \citep{Aug++10}.
As this quantity can be readily predicted with hydrodynamical simulations, our measurement offers one of the most direct ways of comparing models of galaxy evolution with observations.

The structure of this paper is as follows. In \Sref{sc:Strategy} we explain the strategy that we adopted for our measurement. In \Sref{sc:Data} we present the sample and the data used for our study. In \Sref{sc:Photoz} we describe the procedure that we used to obtain redshift information for the lens galaxies. In \Sref{sc:Model} we introduce the model used for our analysis. Section~\ref{sc:Results} shows the results. We discuss our modelling choices in \Sref{sc:Disc} and conclude in \Sref{sc:Concl}.
Throughout our paper, we assume a flat $\Lambda$CDM cosmology model, with $\Omega_{\rm m}=0.3$. We keep the dependence on the value of the Hubble constant explicit through the parameter $h=H_0/(100\,\mathrm{km}\,\mathrm{s}^{-1}\,\mathrm{Mpc}^{-1})$. 

\section{\label{sc:Strategy} Strategy}

Our goal is to infer the properties of the general population of elliptical galaxies, using observations from a sample of strong lenses.
The problem can be summarised as follows.
Each strong lens consists of a foreground galaxy, described with a set of parameters $\psilens$, and a background source, described with a different set of parameters $\psisource$. 
Foreground galaxies and background sources are drawn from populations described by probability distributions $\prlens(\psilens)$ and $\prsource(\psisource)$, respectively. 
With these definitions, the distribution of the population of strong lenses is given by \citep{Son22a}
\begin{equation}\label{eq:one}
\prsl(\psilens,\psisource) \propto \prlens(\psilens)\,\prsource(\psisource)\,\psel(\psilens,\psisource)\,.
\end{equation}
In the above equation, $\psel(\psilens,\psisource)$ is the lens selection probability, that is, the probability of a foreground-background galaxy pair with parameters $(\psilens,\psisource)$ to produce a strong lens that is included in our lens sample.
Our primary target is $\prlens(\psilens)$. In order to obtain it, we must invert \Eref{eq:one} above, as the strong lensing data only constrains the left-hand side term, $\prsl$, directly. 
Doing so requires good knowledge of the background source distribution before lensing, $\prsource$, and the selection function, $\psel$.
To obtain the former, we used a prior from the literature (described in \Sref{ssec:sourcepop}). The latter depends critically on the procedure used to define the strong lens sample. To make our analysis as robust as possible, we made the choice of restricting our study to a subset of \Euclid lenses with a well-understood selection probability, so as to simplify the evaluation of the $\psel$ term. 

The foreground galaxy parameters $\psilens$, which are described in detail in \Sref{ssec:foregroundpop}, include quantities related to the stellar component, such as luminosity, size, and redshift. Measuring how elliptical galaxies are distributed in the parameter space defined by these properties does not require strong lensing data. Therefore, we complemented our strong lens sample with a larger sample of non-lens elliptical galaxies, in order to get a firmer handle on the $\prlens(\psilens)$ distribution. We also did this because the difference in the properties of lenses and non-lenses encodes information on the mass distribution of the galaxy population \citep{Son++23}.

The key ingredient in our analysis is the mass model of the galaxy population. For the sake of simplicity, we chose a singular isothermal sphere (SIS) model, with a three-dimensional density profile $\rho(r)\propto r^{-2}$.
Although this is a simpler model compared to many strong lensing studies from the literature, we verified that our results do not change under reasonable deviations from isothermality (\Sref{sc:Disc}).

The main observable that we used to constrain the model is the Einstein radius, $\tein$, which is approximately the radius of the circle that encloses the multiple images of a lensed source. For each lens, we measured $\tein$ directly from the \Euclid images. The Einstein radius can be related to the Einstein mass $\mein$, which is the total projected mass enclosed within an aperture of circularised radius $\tein$ centred on a lens,
\begin{equation}\label{eq:mein}
\mein = \pi(\tein\,D_{\mathrm{g}})^2\,\Sigma_{\mathrm{cr}}(\zlens,\zsource)\,.
\end{equation}
In the above equation, $D_{\mathrm{g}}$ is the angular diameter distance of the lens\footnote{We use the subscript 'g', instead of the more traditional 'd', as we are referring to all foreground galaxies, not only deflectors in actual strong lens systems.}, and $\Sigma_{\mathrm{cr}}$ is the critical surface mass density for lensing. This is defined as
\begin{equation}
\Sigma_{\mathrm{cr}} = \frac{c^2}{4\pi G_\sfont{N}}\frac{D_\mathrm{s}}{D_{\mathrm{g}}\,D_{\mathrm{gs}}}\,,
\end{equation}
where $D_{\mathrm{s}}$ and $D_{\mathrm{gs}}$ are the angular diameter distances of the source with respect to the observer and to the lens, respectively. 
Therefore, in order to convert measurements of $\tein$ into measurements of mass, information on the redshift of the lens and the source is needed. We only have spectroscopic redshifts for approximately $20\%$ of the lens galaxies, and only three of the lensed sources in our sample. 
Although spectroscopic follow-up observations of \Euclid lenses have been planned, we do not realistically expect to obtain a complete spectroscopic coverage of the full \Euclid strong lens sample in the near future.
Therefore, we opted for a solution that allowed us to take advantage of all available strong lensing data, regardless of spectroscopy.

To obtain redshift information for the lens galaxies, 
we chose to leverage the non-lens elliptical galaxy population, for which a much larger number of spectra is available. In particular, we used a spectroscopic sample of elliptical galaxies to infer the mapping between \Euclid photometry and redshift, which we then used as a prior for the lens sample. 

To obtain source redshifts, we adopted an approach called photogeometric redshift, first introduced by \citet{Ruf++11} and later refined by \citet{Son22b}. The idea is to simultaneously infer the redshift of the source and the mass of the lens, while taking into account how the lensing cross-section varies as a function of these two properties. In this approach, the flux of the source carries important information, as brighter galaxies tend to be preferentially located at a lower redshift. We used a prior on the distribution of galaxies in redshift and flux space to exploit this information.

\section{\label{sc:Data} Data}

Our analysis rests mainly on \Euclid photometric data for a sample of elliptical galaxy strong lenses.
We describe this sample and their data in \Sref{ssec:lenses}. At the same time, we also studied a sample of non-lens elliptical galaxies to better understand those properties of the galaxy population that are directly observable without lensing, including redshift. We describe this sample of non-lenses in \Sref{ssec:nonlens}.
All of the \Euclid data and data products used in this work are taken from Q1. 
We used images in the band \IE of the VIS instrument \citep{EuclidSkyVIS}, and in bands \YE, \JE, and \HE of NISP \citep{EuclidSkyNISP}.

\subsection{\label{ssec:lenses} Lenses}

We built the sample of lenses in such a way as to prioritise the reproducibility of its selection function.
We did so by selecting strong lenses in a region of data space where we are reasonably confident that: 1) the \Euclid lens sample is close to complete; 2) the lensing nature of the objects in this region can be determined with no ambiguity from \Euclid photometric data alone; and 3) the Einstein radius of these lenses can be determined with certainty. In short, we want to ensure completeness and purity of our sample, and robustness of our measurements with respect to modelling choices.
We refer to the subset of systems that satisfy these requirements as the `golden lenses'.
The starting point of our selection of the golden lenses is the ensemble of \Euclid strong lenses located in the region covered by Q1. This ensemble consists of approximately $700$ lens candidates with visual inspection grade A or B, discovered in the searches described in \citet[][W26 from here on]{SLDEA}, \citet{SLDEB}, \citet{SLDEC}, \citet{SLDEF}, \citet{Ber++26}, and \citet{Q1-Xu}.
We then applied cuts to this initial sample:
\begin{enumerate}
\item Lens galaxy \IE flux, $\fluxvis$, larger than $5\,\mujansky$ ($22.15$ in AB magnitude), as measured by fitting the lens light with a de Vaucouleurs profile.
\item Lens galaxy \IE axis ratio larger than $0.5$.
\item Isolated: no neighbours within $3\arcsecond$ of the lens.
\item Lensed source images detected with signal-to-noise ratio $\SN>10$, and subtending an angle of at least $150\degree$ with respect to the lens centre.
\item Total lensed source $\SN$ larger than $10$, when measured over the footprint where the source surface brightness exceeds that of the lens.
\item An Einstein radius larger than $\ang{;;0.7}$.
\end{enumerate}
The application of these cuts requires a model for the surface brightness distribution of the lens and the source. We describe the lens modelling process in the latter part of this section.
First, we explain the rationale for the choices listed above.

We applied the first cut to ensure that the lens passed the pre-selection of the search described in \SLDEA, which is the most comprehensive lens search carried out so far in Q1 data. From the MERge processing function \citep[MER;][]{Q1-TP004} catalogue, \SLDEA\ selected objects for which the quantity $\texttt{FLUX\_DETECTION\_TOTAL}$ is larger than $3.63078\,\mujansky$ (22.5 AB magnitude). 
Typically, this flux indicator includes light from the lensed arcs, and is therefore larger than the de Vaucouleurs flux that we obtained after lens modelling. However, we also found lenses with a larger $\fluxvis$. 
With our more conservative limit at $\fluxvis>5\,\mujansky$, only lenses for which the ratio between $\texttt{FLUX\_DETECTION\_TOTAL}$ and $\fluxvis$ is smaller than $0.7262$ would have been missed by \SLDEA. Less than $2\%$ of the lenses that we modelled have a flux ratio smaller than this threshold.

The rationale for the second cut is to eliminate disk galaxies from the sample, since our focus is on elliptical galaxies.
The third cut is to avoid working with systems of multiple galaxies acting as a lens, which complicates the interpretation of lensing measurements.
We applied the fourth cut to select only lenses with unambiguously detected multiple images that bracket the lens galaxy. Using the best-fit lens model image, we first defined a footprint of the lensed source by selecting all regions of contiguous pixels with surface brightness larger than three times the median root mean square fluctuation of the sky background. We measured the $\SN$ of each separate region, and discarded those with $\SN<10$. 
Finally, we measured the angle $\Delta\phi$ subtended by all pixels associated with regions with $\SN>10$, with respect to the lens centre, and applied the cut $\Delta\phi>150\degree$. The choice of $150\degree$ for the minimum subtended angle is arbitrary, but its exact value has little impact on the sample selection, because the distribution in $\Delta\phi$ of our lens candidates is bimodal, with $\Delta\phi$ nearing $360^\circ$ for most of the golden lenses. 

\begin{figure*}[htbp!]
  \begin{center}
    \includegraphics[width=\textwidth]{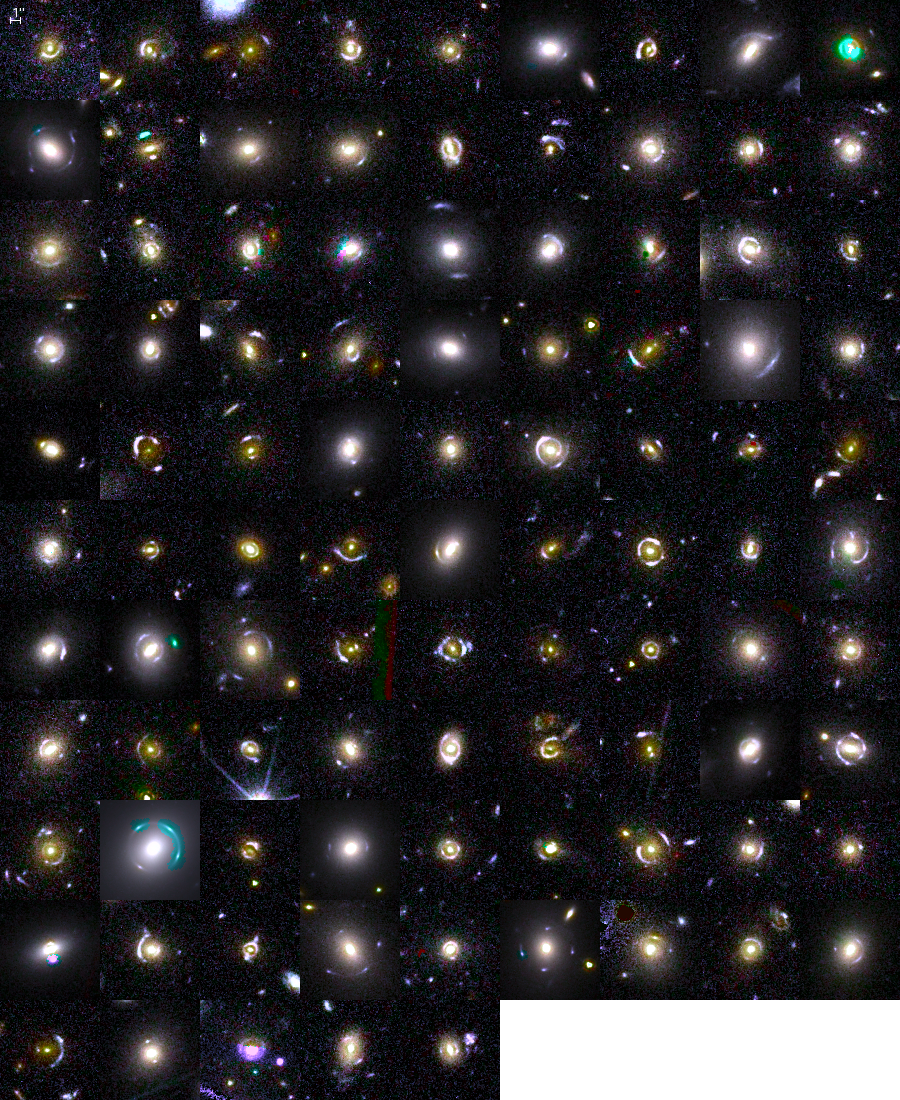}
   \end{center}
\caption{Colour-composite images of the $\Ngold$ golden lenses used in our study. Data from the \IE, \YE, and \JE bands are used for the blue, green, and red channels of the images, respectively. Cyan regions, if present, are due to masked pixels in the NISP images. This can be a severe issue for some of the lensed arcs. Nevertheless, NISP data are only used to characterise the colours of the lens galaxy, which are well measured despite the masking of the source light. Each image is $10''$ on the side.}

\label{fig:golden}
\end{figure*}

The fifth cut selects against lenses for which the contrast between source and lens light is low. These are lenses that are difficult to identify visually, so this cut is needed to maximise completeness. 
Finally, the sixth cut was needed to deal with the fact that it is very difficult to distinguish real lenses from false positives, such as ring galaxies, when the Einstein radius is small. 
Although removing low-$\tein$ lenses from the sample leads to a larger spread between the properties of lenses and non-lens galaxies \citep{Son++23}, this is not a problem for our analysis, because we forward model all of our selection cuts, including that on the Einstein radius.
The final sample consists of $\Ngold$ golden lenses. 
These are high-confidence strong lenses, based on our ability to reproduce the morphology of the lensed arcs with a simple lens model.
Their images are shown in \Fref{fig:golden}.

This number is about a factor of seven smaller than the initial sample of grade A or B lenses in Q1. The reasons for such a sharp compression are various. A few tens of lens candidates were eliminated at the catalogue level, based on their lens galaxy being too faint, bringing the sample down to $693$ candidates. Of these, 119 were found not to show evidence of multiple imaging, in the sense of our golden lens definition. These objects are likely gravitational lenses, but do not have multiple images on opposite sides of the lens galaxy, and therefore do not pass our cuts. A further $97$ were eliminated due to the presence of neighbours, and $64$ lenses were discarded due to the lens galaxy having a clear disk morphology (most of which would likely not have passed the second cut in our golden lens definition, had they been modelled). The remaining $\Nmodelled$ lens candidates were fitted with a lens model (described later in this section).
Of them, $47$ were discarded for the lens being too faint, after being fit with a de Vaucouleurs profile. Then, a further $190$ were found not to show evidence of multiple images with sufficiently high $\SN$. That is, they failed to meet the requirements of point $4$ in the above list of cuts.
Of the remaining ones, $45$ were discarded for their Einstein radius being too small ($14$ of them cannot be distinguished from ring galaxies), one for the lens being too disky, and $13$ for the source-to-lens contrast being too low (criterion $5$ in the above list). 
Finally, we removed $\Namb$ candidates from the remaining $112$, because we deemed them likely not to be lenses, based on our difficulty in qualitatively matching the supposed arc features with a lens model. We discuss these and other ambiguous lens candidates in Appendix~\ref{sect:ambiguous}. Although we cannot exclude with certainty that some of these systems are actual lenses, we verified that the results of our analysis are robust with respect to small changes to the lens sample.

Of the $\Ngold$ golden lenses, $\Nspecz$ have spectroscopic redshifts. Twelve of these redshifts come from the Dark Energy Spectroscopic Instrument Data Release 1 \citep[DESI DR1;][]{DESI1}, five from the Australian Dark Energy Survey \citep[OzDES;][]{OzDES}, and one from the PRIsm MUlti-object Survey \citep[PRIMUS;][]{Coi++11}. We have spectroscopic redshifts of only three lensed sources: EUCL\,J175107.51+653150.6 at $\zsource=0.406$ \citep[NGC\,6505;][]{ORiordan25}, EUCL\,J180354.65+643421.6 at $\zsource=1.897$, and EUCL\,J175049.89+665454.5 at $\zsource=1.956$ \citep{SLDEB}.

We fitted $\Nmodelled$ lens candidates with a lens model.
The main goals of this modelling procedure are to measure the Einstein radius of the lenses, to provide a description of the surface brightness distribution of the lens and the background source, and to measure the colours of the lens galaxy.
The data used for modelling were coadded images, as well as noise maps and models of the point-spread function (PSF), as produced by MER. This PSF is based on stacked \texttt{PSFEx} models \citep{Ber11}, sampled at $\ang{;;0.1}/\mathrm{pixel}$, and not accounting for wavelength dependence. We obtained these data from ESA Datalabs \citep{Datalabscite}.

We carried out the modelling process by first fitting a PSF-convolved S\'{e}rsic model to the \IE lens light distribution, while masking out lensed arcs and neighbouring objects.
We used the object detection software \texttt{SExtractor} \citep{B+A96} for this purpose.
We visually inspected the lens-subtracted images, along with the \texttt{SExtractor} segmentation maps, and classified the detected regions into lensed source images and foreground contaminants.
We used the former to obtain a starting guess for the lens model, and masked out the latter.
Then, we added an exponential profile to describe the background source, and a singular isothermal ellipsoid (SIE) to model the mass of the lens.
We imposed the centre of mass to be the same as the centre of light.
We used the lens modelling software \texttt{Glafic} \citep{Ogu10} to predict the surface brightness distribution of the lensed source, and convolved it with the PSF to compare it with the \IE data.
Finally, we optimised for the lens and source model parameters.
For most systems in the sample, the model obtained after this optimisation was sufficiently good for our purposes, in the sense that it reproduced the positions and shapes of all of the observed lensed images, so that the Einstein radius was well constrained. For $\Nmanual$ lenses, however, we 
manually added elements of complexity to the model, such as a second S\'{e}rsic component for the lens light, external shear, or a spatial offset between the lens light and mass.
A few of the lenses in our sample have two lensed sources at two different redshifts, as shown by \citet{SLDED}. In those cases, we only modelled the arcs produced by the lowest-redshift source (i.e. the innermost arcs), as that is sufficient to measure the Einstein radius.
Once we obtained a satisfactory model, we fixed its structural parameters and optimised solely for the amplitude of each light component in the three NISP bands.
In parallel to this last step, we also fitted a model with a de Vaucouleurs profile (i.e. a S\'{e}rsic profile with index $n=4$) for the lens light, optimising its structural parameters on the \IE image while keeping the mass and source parameters fixed. 
The motivation for this additional fit is that our science analysis is based on a de Vaucouleurs model for the lens, which we prefer over the more general S\'{e}rsic in virtue of its reduced number of degrees of freedom (two, instead of three).

\begin{figure*}[htbp!]
  \centering
    \includegraphics[width=\textwidth]{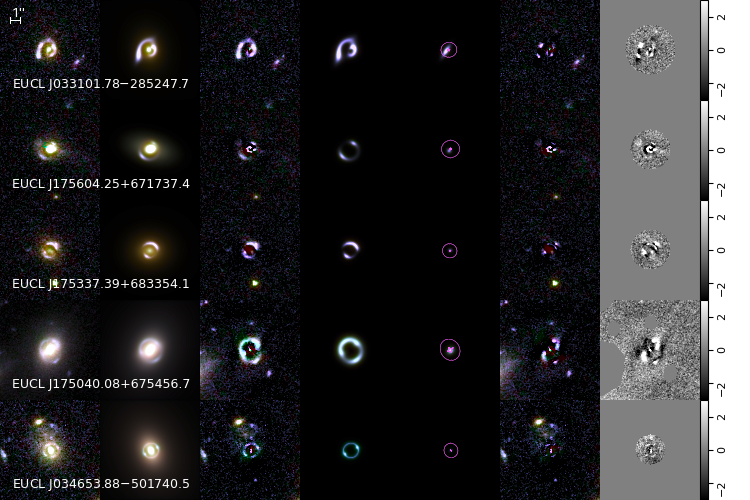}
\caption{Example lens models. From left to right, each column shows the original image, the full model, the lens-subtracted image, the source-only model, the original (before lensing) source-plane image along with the caustic lines of the lens, the residual image, and the VIS-band residual normalised by the uncertainty. All colour images have the same brightness and colour scale. Although colour-composite images are shown here, only the \IE data were used to optimise the lens model parameters. In the last panel, masked pixels are set to zero. Each image is $10''$ on the side.}
\label{fig:modelexample}
\end{figure*}

Figure~\ref{fig:modelexample} shows examples of models of a few lenses.
Although in several cases our model is unable to reproduce small-scale features in the images of the lensed source, it can qualitatively match the positions, sizes, shapes, and orientations of the multiple images. 

\begin{figure*}
\includegraphics[width=\textwidth]{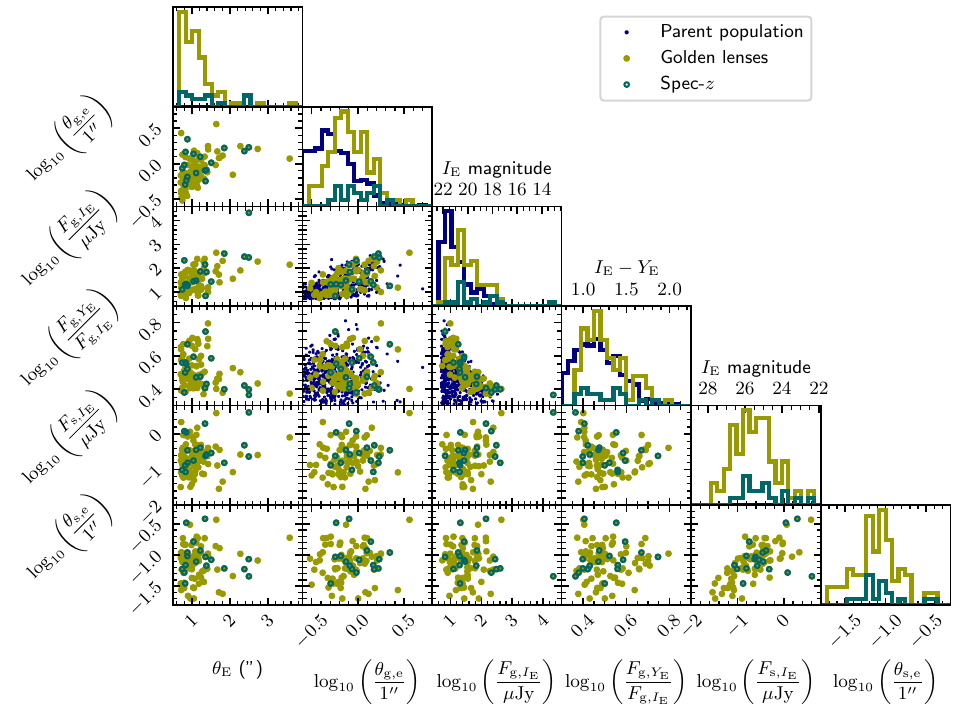}
\caption{Distribution of observed properties of the lenses. These are the Einstein radius, flux, half-light radius, and \YE- to \IE-band flux ratio of the lenses, and the unlensed flux and half-light radius of the background sources. 
The distribution of the non-lens elliptical galaxy sample (i.e. the parent population) is also shown.
Scatter points of lenses with spectroscopic redshift are circled in green.
The histograms of the parent and lens population are normalised to one. The latter include the lenses with spec-$z$. The spec-$z$ sample histogram is normalised to the fraction of lenses with spectroscopic redshift.
AB magnitude scales are shown in correspondence to the flux and colour columns.
}
\label{fig:obs}
\end{figure*}

Figure~\ref{fig:obs} shows the distribution in Einstein radius, half-light radius, lens flux, and ratio between \YE- and \IE-band flux of the lenses. 
We see a clear correlation between flux and \YE--\IE flux ratio. This reflects the underlying correlation between colour and redshift, which we exploit to gain redshift information, as explained in \Sref{sc:Photoz}.
Figure~\ref{fig:obs} also shows the distribution in flux and angular half-light radius of the background sources, before lensing. We see a clear correlation between the two. Although we do expect brighter sources to be, on average, more extended, part of this correlation could also be due to a degeneracy between these two quantities. This is because neither the total flux nor the half-light radius is directly observable, but is derived under the assumption of an exponential surface brightness profile, often extrapolated to regions that are not constrained by the data.
Table~\ref{tab:lenses} lists a selection of properties of the golden lenses. A more complete list with all of the parameters of the model is available online\footnote{\url{https://zenodo.org/records/21345571}}. We also provide individual lens models in the code-independent organised lens standard \citep[COOLEST;][]{Gal++23}.

\begin{table*}
\caption{Selected observables properties of the golden lenses. 
}
\smallskip
\label{tab:lenses}
\smallskip
\begin{tabular}{lccccccc}
Lens name & $\zlens$ & $z_{\mathrm{g,phot}}$ & $\logten{(\lensflux/\mujansky)}$ & $\teff$ $(\arcsecond)$ & $\tein$ $(\arcsecond)$ & $\zsource$ & $\logten{(\sourceflux/\mujansky)}$ \\
\hline
\hline
EUCL\,J032408.71$-$271429.5 & $\ldots$ & $0.75\pm0.05$ & $1.38$ & $1.35$ & $0.75$ & $\ldots$ & $-0.47$ \\
EUCL\,J032618.89$-$281123.0 & $0.775$ & $0.72\pm0.05$ & $1.10$ & $0.86$ & $0.75$ & $\ldots$ & $-1.01$ \\
EUCL\,J032723.88$-$263653.0 & $\ldots$ & $0.97\pm0.05$ & $1.19$ & $1.59$ & $1.43$ & $\ldots$ & $-0.84$ \\
EUCL\,J032751.20$-$272533.4 & $\ldots$ & $0.63\pm0.04$ & $1.31$ & $0.75$ & $0.80$ & $\ldots$ & $-0.51$ \\
EUCL\,J032923.51$-$290907.6 & $0.730$ & $0.71\pm0.05$ & $1.29$ & $1.47$ & $0.83$ & $\ldots$ & $-1.06$ \\
EUCL\,J032929.39$-$262115.1 & $\ldots$ & $0.26\pm0.03$ & $2.28$ & $0.97$ & $1.31$ & $\ldots$ & $-0.77$ \\
EUCL\,J033101.78$-$285247.7 & $\ldots$ & $0.93\pm0.05$ & $0.87$ & $0.51$ & $0.77$ & $\ldots$ & $-0.03$ \\
EUCL\,J033410.71$-$281210.1 & $0.289$ & $0.33\pm0.04$ & $2.15$ & $0.80$ & $1.35$ & $\ldots$ & $0.29$ \\
EUCL\,J033414.91$-$280046.4 & $\ldots$ & $0.40\pm0.05$ & $1.92$ & $0.96$ & $0.89$ & $\ldots$ & $-0.07$ \\
EUCL\,J033459.29$-$263431.6 & $0.346$ & $0.32\pm0.03$ & $2.45$ & $1.70$ & $2.49$ & $\ldots$ & $-0.63$ \\
\end{tabular}
\tablefoot{Spectroscopic redshift $\zlens$ (where available), photometric redshift $z_{\mathrm{g,phot}}$ (from the procedure described in \Sref{sc:Photoz}), \IE flux, half-light radius $\teff$, and Einstein radius $\tein$ of the lens galaxies; \IE flux and spectroscopic redshift (where available) of the lensed source. All fluxes are corrected by Galactic extinction, using the \citet{Fit99} dust extinction curve. Lens fluxes and sizes were obtained by fitting a de Vaucouleurs profile to the \IE image. 
Only the first ten lenses are shown in print. The full table is available online.}
\end{table*}

We estimated the completeness of the golden lens sample to be $64\%$. Appendix~\ref{sect:compl} explains how we obtained this value.
For the rest of this paper, we assume that our sample is effectively complete, meaning that, if any lenses are missing, they are indistinguishable from the rest of the sample.

\subsection{\label{ssec:nonlens} Non-lens sample}

The purpose of the non-lens sample of elliptical galaxies is to provide prior information on the properties of the elliptical galaxies population, to be used in the strong lensing analysis. 
For this reason, it is important that this sample is built with the same selection steps that we applied to the golden lenses (where applicable), and that their photometric measurements are carried out consistently with our lens models.

To select a representative sample of non-lens elliptical galaxies, we drew random positions over the Q1 area, and searched for galaxies lying within $5\arcsecond$ of each drawn position. 
From the MER catalogue, we first selected all objects with $\texttt{FLUX\_DETECTION\_TOTAL} > 3.63078\,\mujansky$ (i.e. the same cut used for the pre-selection of lenses in \SLDEA), $\texttt{SERSIC\_SERSIC\_VIS\_AXIS\_RATIO} > 0.4$ (we applied a more stringent cut on axis ratio in a second step, described below), and $\texttt{POINT\_LIKE\_PROB} < 0.2$. We also applied a lower limit to the \YE- to \IE-band flux ratio, $\logten{(F_{\mathrm{g},\YE}/\lensflux)} > 0.3$, using aperture fluxes within one FWHM in PSF-matched images. The reason for this colour cut is that the golden lenses are all observed to be redder than this limit (\Fref{fig:obs}).
With this procedure, we searched over an effective area of $0.19$~deg$^2$, and found $1549$ galaxies (which extrapolate to $\sim5\times10^5$ over the entire Q1 area). We visually inspected them to select galaxies with elliptical morphology, and no neighbours within a $3\arcsecond$ aperture (this is the same isolation criterion that we adopted for the lenses), reducing the sample size to $657$. Finally, we fitted the \IE image of each galaxy with a de Vaucouleurs model, and selected all galaxies with $\fluxvis>5\,\mujansky$ and axis ratio larger than $0.5$. 

The final sample consists of $534$ non-lens elliptical galaxies. Of them, $34$ have spectroscopic redshifts. We measured their NISP fluxes via forced photometry\footnote{Forced photometry means that we fixed the structural parameters of the de Vaucouleurs surface brightness model (centroid, half-light radius, etc.) to be the same in all bands.} to the best-fit \IE de Vaucouleurs model. Fluxes, half-light radii, and \YE- to \IE-band flux ratios of these galaxies are shown in \Fref{fig:obs} alongside the lens sample.
We can see that the two populations overlap, with the non-lenses extending to a region of low flux and relatively blue colour that is not covered by the lenses. Those are generally low-redshift and low-mass galaxies, with a very small cross-section for strong lensing.

\section{\label{sc:Photoz} Photometric redshifts and physical properties}

Our basis for obtaining photo-$z$s is an empirical relation between redshift and the $(\IE,\YE,\JE,\HE)$ fluxes of elliptical galaxies, which we derived as follows.
We selected a subset of $1000$ galaxies from the sample of objects with DESI DR1 spectra that lie in the Q1 area. When drawing this subset, we gave a larger weight to the brightest galaxies at each redshift, so as to obtain a sample that is broadly representative of strong lens galaxies. We visually inspected this sample, and selected $252$ isolated elliptical galaxies. Then, we fitted a de Vaucouleurs profile to their \IE image, and obtained NISP band fluxes via forced photometry.

\begin{figure}
\includegraphics[width=\columnwidth]{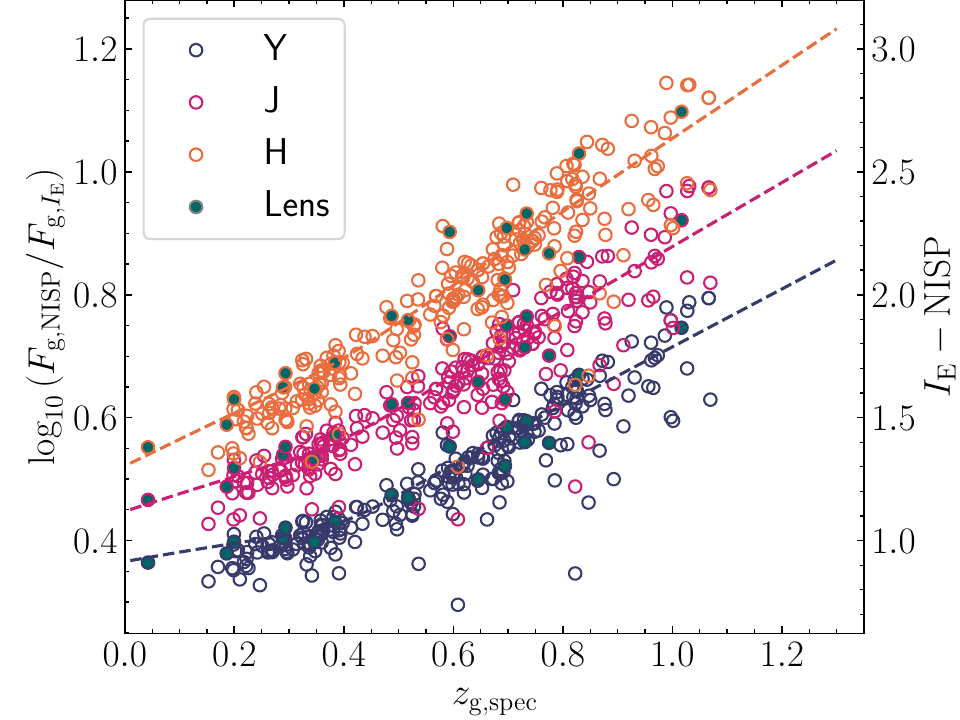}
\caption{
NISP to \IE flux ratio of elliptical galaxies as a function of redshift. Empty circles: $252$ elliptical galaxies with DESI DR1 spectroscopy. Filled circles: $\Nspecz$ golden lenses (not used in the inference of the photo-$z$).
Dashed lines indicate the model $\mu_{\lambda}(\zlens,\fluxvis)$, evaluated at $\fluxvis=F_*(\zlens)$, that we inferred.
The right axis indicates colours in units of AB magnitude.
\label{fig:colourvsz}
}
\end{figure}

Figure~\ref{fig:colourvsz} shows the ratio between the flux in each NISP band and the \IE flux, as a function of redshift, of these elliptical galaxies with DESI spectra. We can see a clear correlation for all three bands. 
In light of these data, we fitted the following model to the distribution in the four-band flux of elliptical galaxies as a function of redshift,
\begin{equation}\label{eq:photoz}
P(\fluxset|z) = \prod_{\lambda\in\mathrm{NISP}}\mathcal{N}_{\logten{(F_{\mathrm{l},\lambda}/\fluxvis)}}[\mu_\lambda(\zlens,\fluxvis), \sigma_\lambda^2]\,,
\end{equation}
where the product is over the three NISP bands, and the notation $\mathcal{N}_x[\mu,\sigma^2]$ indicates a Gaussian distribution in the variable $x$ with mean $\mu$ and variance $\sigma^2$.
The mean of each Gaussian takes the following form,
\begin{align}
\mu_\lambda(\zlens,\fluxvis) & = \mu_{\lambda,0} + \zeta_{\lambda,1}(\zlens - \zt)\,\{1 - s[b_z(\zlens-\zt)]\} \nonumber \\
& + \zeta_{\lambda,2}(\zlens-\zt)\,s[b_z(\zlens-\zt)] + \omega\logten{\left(\frac{\fluxvis}{F_*(\zlens)}\right)},\label{eq:meanfluxrat}
\end{align}
where $s$ is a sigmoid function,
\begin{equation}\label{eq:sigmoid}
s(x) = \frac{1}{1 + \mathrm{e}^{-x}}\,,
\end{equation}
$F_*$ is a redshift-dependent characteristic flux,
\begin{equation}
F_*(\zlens) = \left[\frac{D_{\mathrm{L}}(0.5)}{D_{\mathrm{L}}(\zlens)}\right]^2 100\,\mujansky\,,
\end{equation}
and $D_{\mathrm{L}}$ is the luminosity distance.
In words, we modelled the ratio between the near-infrared and optical flux, in each NISP band, as a Gaussian, with a mean that scales with redshift as a two-piece linear function, with an additional dependence on the \IE flux. The parameter $\zt$ indicates the redshift at which the slope of the colour-redshift relation changes, while $b_z$ governs the smoothness of the transition. 
In all, the model parameters are: $\mu_{\lambda,0}$, $\zeta_{\lambda,1}$, $\zeta_{\lambda,2}$, $\omega_{\lambda}$, and $\sigma_\lambda$, for each NISP filter; $\zt$, and $b_z$. Their inferred values are listed in Table~\ref{tab:photoz}. Throughout the rest of our analysis, we fix the value of each of these parameters to the median of its marginal posterior probability. In \Fref{fig:colourvsz} we show, for each NISP band, the mean colour $\mu_{\lambda}(\zlens,\fluxvis)$, evaluated at the reference flux $F_*(\zlens)$.

\begin{table*}
\caption{Parameters of the photometric redshift model.}
\smallskip
\label{tab:photoz}
\smallskip
\begin{tabular}{cccp{7.5cm}}
Parameter(s) & Posterior ($68\%$) & Eq. Ref. & Description \\
\hline
\hline
$z_t$ & $0.33\pm0.10$ & (\ref{eq:meanfluxrat}) & Transition redshift of the redshift-colour relation.\\
$a_z$ & $6.4\pm8.2$ & (\ref{eq:meanfluxrat}) & Rapidity of the transition from the low- to the high-redshift regime.\\
$\mu_{Y,0},\,\mu_{J,0},\,\mu_{H,0}$ & $0.41\pm0.03,\,0.55\pm0.04,\,0.67\pm0.05$ & (\ref{eq:meanfluxrat}) & Average VIS$-$NISP colour at $z=z_t$, for each NISP band.\\
$\zeta_{Y,1},\,\zeta_{J,1},\,\zeta_{H,1}$ & $0.07\pm0.23,\,0.21\pm0.21,\,0.42\pm0.16$ & (\ref{eq:meanfluxrat}) & Scaling of colour with redshift, at the low-$z$ end.\\
$\zeta_{Y,2},\,\zeta_{J,2},\,\zeta_{H,2}$ & $0.46\pm0.03,\,0.51\pm0.03,\,0.59\pm0.03$ & (\ref{eq:meanfluxrat}) & Scaling of colour with redshift, at the high-$z$ end.\\
$\omega_Y,\,\omega_J,\,\omega_H$ & $0.035\pm0.011,\,0.040\pm0.015,\,0.047\pm0.018$ & (\ref{eq:meanfluxrat}) & Scaling of colour with \IE flux.\\
$\sigma_Y,\,\sigma_J,\,\sigma_H$ & $0.042\pm0.002,\,0.051\pm0.003,\,0.059\pm0.003$ & (\ref{eq:photoz}) & Scatter in colour at fixed flux and redshift.\\
\end{tabular}
\tablefoot{The second column reports the median and $68\%$ credible region of the marginal posterior probability. The third column indicates the equation where each parameter is first defined.}
\end{table*}

\begin{figure}
\includegraphics[width=\columnwidth]{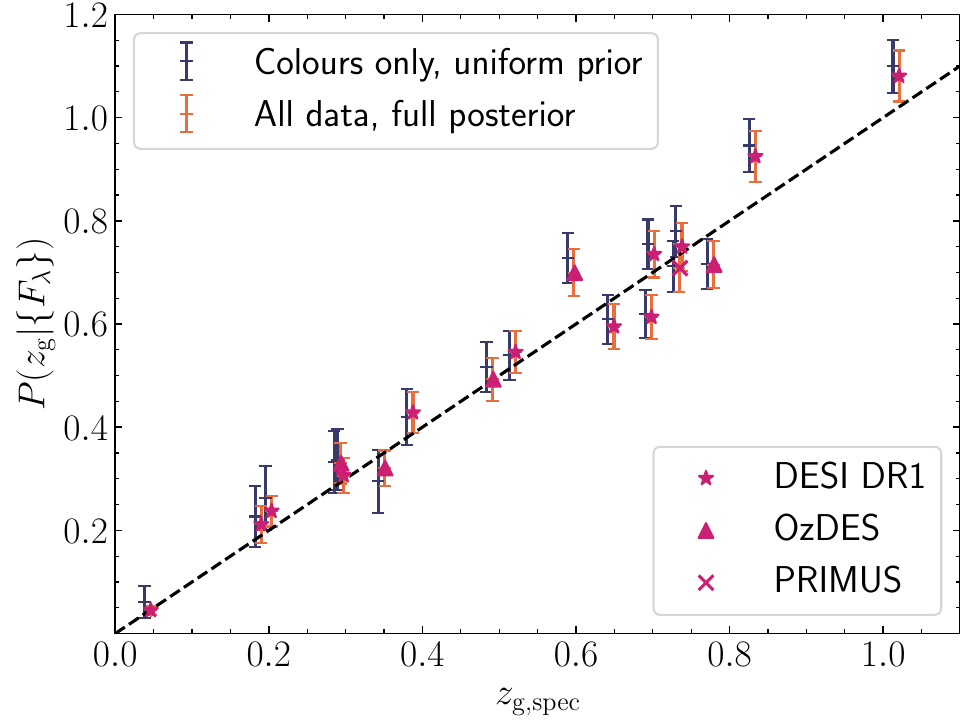}
\caption{Photo-$z$ accuracy test. Posterior predicted redshift as a function of spectroscopic redshift for the lenses with spectroscopic data. Error bars bracket an enclosed probability of $68\%$.
Purple data points were obtained by simply using the model of \Eref{eq:photoz} and assuming a uniform prior on the lens redshift. Orange data points were obtained by using all available information, and our full model of the lens population as a prior.
\label{fig:specztest}
}
\end{figure}

We checked the accuracy of our model with the $\Nspecz$ lenses for which we have a spectroscopic redshift.
For each lens, we computed the posterior probability of the redshift given all available data, which include the lens fluxes, the half-light radius, the Einstein radius, and the intrinsic source flux.
All of these observables can inform the redshift of the lens, when interpreted in the context of a lens model.
The correct prior probability to be used for this purpose is given by the distribution of \Eref{eq:one}, marginalised over the properties of the source, over the mass of the lens, and over the uncertainty of the model. 
Figure \ref{fig:specztest} shows the resulting posterior probability, as a function of the spectroscopic redshift.
There is good agreement between the posterior predicted redshifts and the spectroscopic ones, with no catastrophic outliers.
The average difference between the two is $\Delta z=0.014$, which translates into an error on $\Sigma_{\mathrm{cr}}$ of less than $0.1\%$, much smaller than the precision on our measurement of the average mass-to-light ratio of elliptical galaxies. Therefore, we are not concerned with this bias impacting the results of our work. 

In order to gauge the relative importance of the photo-$z$ model of \Eref{eq:photoz}, in \Fref{fig:specztest} we also show posterior predicted redshifts obtained by assuming a uniform prior on the lens redshift, and ignoring any piece of information other than the colours. A comparison with the full posterior shows that colour information alone can constrain the redshift with a typical precision of $0.05$. This uncertainty is set by the observed scatter in the colour of elliptical galaxies at fixed redshift.

One caveat in this analysis is that the spectroscopic redshifts that we used to constrain our model are all drawn from DESI, which specifically targeted luminous red galaxies by means of cuts in flux--colour space \citep{Zho++23}. These galaxies are not necessarily a representative subsample of the entire population of elliptical galaxies, and therefore the mapping between photometry and redshift that we derived might be biased. As most of the spectroscopic redshifts for our lens sample are also from DESI, this bias might be underestimated in our comparison of \Fref{fig:specztest}. However, luminous red galaxies are fairly representative of the typical \Euclid lens, since, of the lenses that lie in the region of the sky covered by DESI DR1, roughly a third are DESI targets. 
When considering the six lenses with a spectroscopic redshift from a survey other than DESI, we found that their average photo-$z$ is larger than their average redshift by only $0.01$, with a $0.02$ uncertainty. Taking this as an estimate of the bias due to the spectroscopic training sample, we conclude that this bias is not significant in the context of our mass measurement.

Using the photo-$z$, we derived physical properties of the galaxies in our samples.
We obtained luminosities by fitting spectral templates to the observed fluxes, which we corrected for Galactic extinction using the model of \citet{Fit99}.
For the fit, we used elliptical-type templates from the Kinney--Calzetti spectral atlas \citep{Cal++94,Kin++96}, as provided by the software \texttt{LePhare} \citep{Arn++99,Ilb++06}. We complemented these with ultra-red templates, which we built by applying dust attenuation to the {\sc Ell} template. This was made necessary by the fact that, for the reddest galaxies in our sample, none of the standard Kinney-Calzetti templates was able to provide a good fit. 
We normalised the templates to the same rest-frame SDSS $r$-band luminosity, $L_r$, and sorted them by their increasing $L_r/L_g$, where $L_g$ is the rest-frame SDSS $g$-band luminosity. Throughout the rest of the paper, we refer to the ratio $L_r/L_g$ as the rest-frame colour.
Then, for a given redshift, we computed their \Euclid fluxes and interpolated them, to obtain flux as a continuous variable $T$, which spans templates from the bluest to the reddest. Given $\zlens$ and $\fluxset$, we optimised for $T$ and for the normalisation of the template, to obtain luminosity and rest-frame colour.

\begin{figure*}
\includegraphics[width=\textwidth]{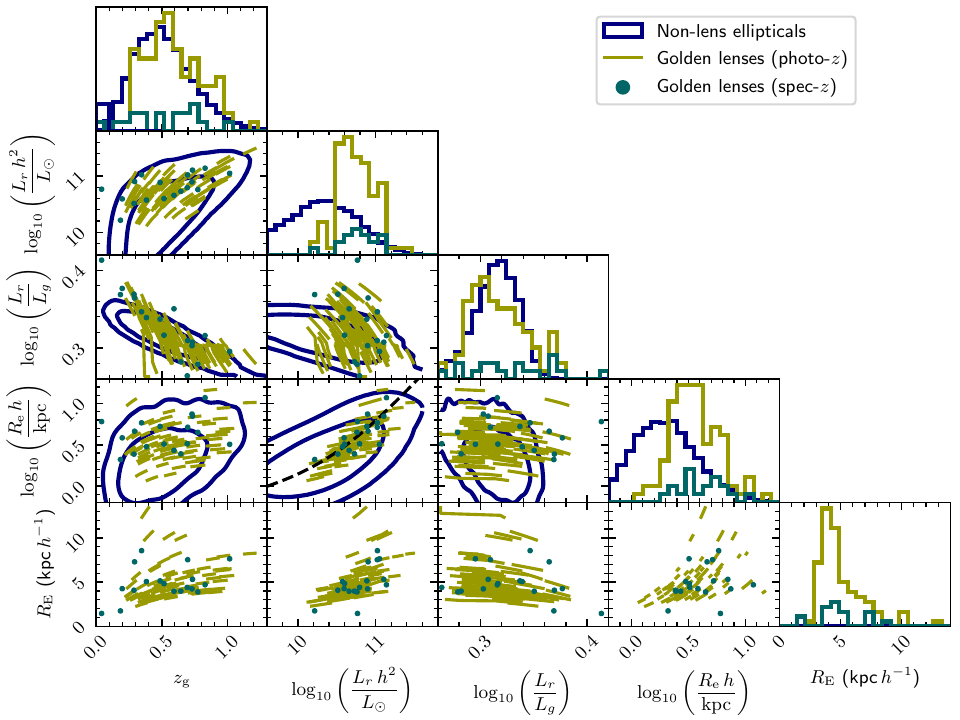}
\caption{
Physical properties of the lenses in our sample. These include redshift, luminosity, rest-frame colour, half-light radius, and Einstein radius. For the lenses with spectroscopic redshift, values are shown with scatter points. For the remaining lenses, the curve corresponding to the $68\%$ credible region of the photo-$z$ is shown. The gold histograms include both spectroscopic and photometric lenses. Contours show the 68\% and 95\% enclosed probability of the distribution of non-lens elliptical galaxies. The dashed black curve shows the luminosity-size relation of non-lens elliptical galaxies, evaluated at $\zlens=0.5$.
\label{fig:physical}
}
\end{figure*}

Figure~\ref{fig:physical} shows physical Einstein radii, half-light radii, luminosities, rest-frame colours, and redshifts, for the lenses in our sample.
The median Einstein radius is close to $\fivekpc$. This motivated our choice of the mass enclosed within that aperture as the main target for our analysis.

\section{\label{sc:Model} Model}

We want to fit a model distribution of the form of \Eref{eq:one} to the ensemble of golden lenses. The main quantities that we want to constrain are a set of parameters describing the population of elliptical galaxies. We indicate them collectively as $\hyperpars$, and we introduce them in \Sref{ssec:foregroundpop}. The posterior probability of $\hyperpars$ is
\begin{equation}\label{eq:posterior}
P(\hyperpars|\data) \propto P(\hyperpars)\,P(\data|\hyperpars)\,,
\end{equation}
where $\data$ is the ensemble of data, $P(\hyperpars)$ is the prior probability on the population parameters, and $P(\data|\hyperpars)$ is the likelihood of observing the data given the model.
The data consist of the Einstein radius, the unlensed source \IE flux, the \IE and NISP lens fluxes, and the lens half-light radius,
\begin{equation}
\data = (\tein,\sourceflux,\fluxset,\teff)\,.
\end{equation}
In so doing, we compress all of the information from the lens model into the pair of parameters $(\tein,\sourceflux)$.
The data for each lens are independent of each other, therefore we can write
\begin{align}
P(\data|\hyperpars) & = \prod_iP(\datai|\hyperpars) \nonumber \\
& = \prod_i\int \diff\psilensi \diff\psisourcei\, P(\datai|\psilensi,\psisourcei)\, \prsl(\psilensi,\psisourcei|\hyperpars) \label{eq:integrals}\,.
\end{align}
The second equality means that the likelihood of the data for the $i$th lens can be obtained while marginalising over its individual parameters.
The probability distribution $\prsl(\psilens,\psisource|\hyperpars)$ is, in general, not analytical. This means that, in order to obtain an accurate result, we have to actively ensure that it is always properly normalised. That is,
\begin{equation}\label{eq:norm}
\int \prlens(\psilens|\hyperpars)\,\prsource(\psisource)\,\psel(\psilens,\psisource)\, \diff\psilens \diff\psisource = 1\,,
\end{equation}
for any value of the parameters $\hyperpars$. Assuming that the background sources are randomly distributed in the sky, the integral over the angular position of the source can be computed separately from that over the other variables. We can therefore introduce the strong lensing cross-section, defined as the following integral over the source position $\vec\beta$,
\begin{equation}
\crosssect(\psilens,\psisourcenobeta) \equiv \int \psel(\psilens,\psisourcenobeta) \,\diff^2\beta\,,
\end{equation}
where $\psisourcenobeta$ is the set of source parameters excluding $\vec\beta$, and the integral extends over the whole sky.
With this definition, \Eref{eq:norm} becomes
\begin{equation}\label{eq:norm2}
\int \prlens(\psilens|\hyperpars)\,\prsource(\psisource)\,\crosssect(\psilens,\psisourcenobeta) \,\diff\psilens \diff\psisourcenobeta = 1\,.
\end{equation}
In the rest of this section, we describe the details of the model.

\subsection{Foreground galaxy population}\label{ssec:foregroundpop}

We describe the population of elliptical galaxies with a probability distribution of the following form,
\begin{equation}\label{eq:fgmodel}
\prlens(\psilens) = \prstar(\zlens,L_r,L_r/L_g,\reff)\,\prmass(\mfive|\zlens,L_r,\reff)\,,
\end{equation}
where $\prstar$ describes the properties of the stellar component, while $\prmass$ describes the mass distribution.
In \Eref{eq:fgmodel}, $\mfive$ is the total projected mass enclosed within an aperture of $\fivekpc$. When comparing our spherical mass distribution with the SIE-based lens models of the golden lenses, we focus exclusively on the Einstein radius, which is a circularly-averaged quantity. Therefore, $\mfive$ uniquely determines the projected mass distribution, which is given by
\begin{equation}
\Sigma(R) = \frac{M_5}{2\pi(\fivekpc)^2}\left(\frac{\fivekpc}{R}\right)\,.
\end{equation}
We parameterise the first factor of \Eref{eq:fgmodel} as
\begin{equation}
\prstar(\zlens,L_r,L_r/L_g,\reff) = P(\zlens)\,P(L_r|\zlens)\,P(L_r/L_g|\zlens)\,P(\reff|\zlens,L_r)\,.
\end{equation}
For the marginal distribution in redshift, $P(\zlens)$, we adopt a generalised gamma distribution,
\begin{equation}\label{eq:gengamma}
P(\zlens) = \frac{|\czpar|}{\Gamma(\azpar)\,\zscalepar}\left(\frac{\zlens}{\zscalepar}\right)^{\azpar\czpar-1}\exp{\left[-\left(\frac{\zlens}{\zscalepar}\right)^{\czpar}\right]}\,.
\end{equation} 
For the luminosity distribution we adopt a Schechter function,
\begin{equation}\label{eq:schech}
P(L_r|\zlens) \propto \left[\frac{L_r}{L_*(\zlens)}\right]^\alpha\mathrm{e}^{-L_r/L_*(\zlens)}\,,
\end{equation}
with a redshift-dependent characteristic luminosity,
\begin{equation}\label{eq:muschech}
\logten{[\frac{L_*(\zlens)\,h^2}{L_{r,\odot}}]} = \logten{\left(\frac{L_{*,0}\,h^2}{L_{r,\odot}}\right)} + \zeta_\mathrm{L}\,(\zlens - 0.5)\,.
\end{equation}
The luminosity distribution is truncated at a minimum value $L_{\mathrm{min}}$, set by our selection cut on the \IE-band flux. The exact value of $L_{\mathrm{min}}$ depends on the redshift and colour of a galaxy.
We modelled the distribution in rest-frame colour as a Gaussian in $\logten{(L_r/L_g)}$,
\begin{equation}\label{eq:gausscolour}
P(L_r/L_g|\zlens) = \mathcal{N}_{\logten{(L_r/L_g)}}[\mu_\mathrm{C}(\zlens), \sigma_\mathrm{C}^2]\,,
\end{equation}
with a redshift-dependent mean defined as
\begin{equation}\label{eq:mucolour}
\mu_\mathrm{C}(\zlens) = \mu_{\mathrm{C},0} + \zeta_\mathrm{C}\,(\zlens - 0.5)\,.
\end{equation}
Finally, we modelled the size distribution as a Gaussian in the logarithm of the half-light radius,
\begin{equation}\label{eq:gausssize}
P(\reff|\zlens,L_r) = \mathcal{N}_{\logten{(\reff\,h/\mathrm{kpc})}}[\mu_\mathrm{R}(\zlens,L_r), \sigma_\mathrm{R}^2]\,.
\end{equation}
The mean of this Gaussian scales with luminosity and redshift,
\begin{align}
\mu_\mathrm{R}(\zlens,L_r) = & \mu_{\mathrm{R},0} + \zeta_\mathrm{R}(\zlens - 0.5) \nonumber \\
& + \rho_{\mathrm{L},1}\logten{(L_r/L_t)}\,\{1 - s[b_\mathrm{L}\logten{(L_r/L_t)}]\} \nonumber \\
& + \rho_{\mathrm{L},2}\logten{(L_r/L_t)}\,s[b_\mathrm{L}\logten{(L_r/L_t)}]\,, \label{eq:musize}
\end{align}
where $s$ is the sigmoid function defined in \Eref{eq:sigmoid}.

We fitted the parameters of the stellar distribution, $\prstar$, to the sample of non-lens galaxies described in \Sref{ssec:nonlens}. 
The best-fit values of the parameters are reported in \Tref{tab:nonlensfit}.
We express $r$-band luminosities in terms of the solar value, which we take to be $L_{r,\odot} = 6.00\times10^{11}\,\mathrm{W}\,\mathrm{Hz}^{-1}$ \citep{Wil++18}. The distribution $\prstar$ is also shown in \Fref{fig:physical} (blue contours).
\begin{table*}
\caption{Parameters of the galaxy distribution in redshift, luminosity, colour, and size.}
\smallskip
\label{tab:nonlensfit}
\smallskip
\begin{tabular}{cccl}
Parameter(s) & Posterior ($68\%$) & Eq. def. & Description \\
\hline
\hline
$z_0$ & $0.32\pm0.13$ & (\ref{eq:gengamma}) & Scale parameter of the marginal distribution in redshift.\\
$a_z$ & $2.4\pm0.8$ & (\ref{eq:gengamma}) & Exponent of the marginal distribution in redshift.\\
$c_z$ & $1.55\pm0.27$ & (\ref{eq:gengamma}) & Shape parameter of the marginal distribution in redshift.\\
$\logten{[L_{*,0}\,h^2/L_{r,\odot}]}$ & $10.55\pm0.04$ & (\ref{eq:muschech}) & Scale of the luminosity function at $z=0.5$.\\
$\zeta_\mathrm{L}$ & $0.33\pm0.16$ & (\ref{eq:muschech}) & Redshift dependence of the luminosity scale parameter.\\
$\alpha$ & $-0.98\pm0.03$ & (\ref{eq:schech}) & Faint-end slope of the luminosity function.\\
$\mu_{\mathrm{C},0}$ & $0.316\pm0.001$ & (\ref{eq:mucolour}) & Average colour at $z=0.5$.\\
$\zeta_\mathrm{C}$ & $-0.078\pm0.003$ & (\ref{eq:mucolour}) & Redshift dependence of the average colour.\\
$\sigma_\mathrm{C}$ & $0.0101\pm0.0004$ & (\ref{eq:gausscolour}) & Intrinsic scatter in colour, at fixed redshift.\\
$\mu_{\mathrm{R},0}$ & $0.40\pm0.18$ & (\ref{eq:musize}) & Average size at $z=0.5$ and $L_r=L_t$.\\
$\zeta_\mathrm{R}$ & $-0.36\pm0.06$ & (\ref{eq:musize}) & Scaling of the average size with redshift.\\
$\rho_{\mathrm{L},1}$ & $0.34\pm0.09$ & (\ref{eq:musize}) & Faint-end slope of the luminosity-size relation.\\
$\rho_{\mathrm{L},2}$ & $0.93\pm0.21$ & (\ref{eq:musize}) & Bright-end slope of the luminosity-size relation.\\
$\logten{[L_t\,h^2/L_{r,\odot}]}$ & $10.48\pm0.28$ & (\ref{eq:musize}) & Transition luminosity of the luminosity-size relation\\
$b_\mathrm{L}$ & $0.20\pm0.26$ & (\ref{eq:musize}) & Rapidity of the transition of the luminosity-size relation.\\
$\sigma_\mathrm{R}$ & $0.187\pm0.006$ & (\ref{eq:gausssize}) & Intrinsic scatter in size, at fixed redshift and luminosity.\\
\end{tabular}
\tablefoot{The second column reports the median and $68\%$ credible interval of the marginal posterior probability. The third column indicates the equation where each parameter is first defined.}
\end{table*}

We describe the distribution in $\mfive$ of the population of elliptical galaxies as a Gaussian in $\logten{(\mfive\,h/M_\odot)}$, where the factor $h$ follows from the fact that, for a given angular Einstein radius, the Einstein mass is inversely proportional to the Hubble constant. The model is
\begin{equation}\label{eq:mfive}
\prmass(\mfive|\zlens,L_r,\reff) = \mathcal{N}_{\logten{(\mfive\,h/M_\odot)}}[\mu_\mathrm{M}(\zlens,L_r,\reff), \sigma_\mathrm{M}^2]\,,
\end{equation}
with mean
\begin{align}
\mu_\mathrm{M}(\zlens,L_r,\reff) = \,& \mu_{\mathrm{M},0} + \beta_z\,(\zlens - 0.5) \nonumber \\
& + \beta_\mathrm{L}\,[\logten{(L_r\,h^2/L_{r,\odot})} - 10.7] \nonumber \\
& + \beta_\mathrm{R}\,[\logten{(\reff\,h/\mathrm{kpc})} - 0.5] \label{eq:mumfive}\,.
\end{align}
In summary, the set of model parameters that we wish to infer from the \Euclid strong lensing data is
\begin{equation}\label{eq:hyperpars}
\hyperpars = (\mu_{\mathrm{M},0},\beta_z,\beta_\mathrm{L},\beta_\mathrm{R},\sigma_\mathrm{M})\,.
\end{equation}
This is the simplest model that allows us to reach the main goal of our work, which is to determine the average $\mfive$ of the population of elliptical galaxies, and its scaling with redshift, luminosity, and size.
We consider the model satisfactory if it can reproduce the observed distribution in Einstein radius of our lens sample. We verify whether this requirement is met with posterior predictive tests, as explained in \Sref{sc:Results}.

\subsection{Background source population}\label{ssec:sourcepop}

\begin{figure}
\includegraphics[width=\columnwidth]{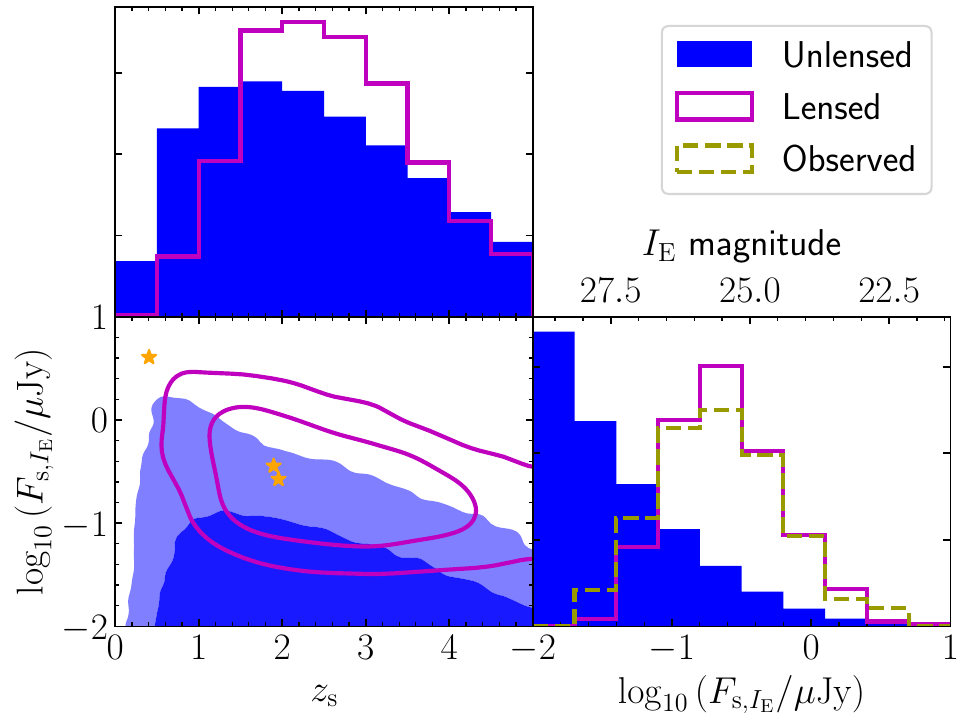}
\caption{
Prior probability distribution of the background source population, in redshift and \IE flux space. Filled contours: distribution before lensing \citep[from][]{Par++16}. Solid lines: lensed distribution. Dashed histogram (bottom-right panel): observed distribution in source flux. The three stars indicate the three lensed sources for which we have a spectroscopic redshift. The lensed distribution contours are obtained using the best-fit model parameters, and therefore do not take into account the uncertainty on the lens population properties.
\label{fig:sourceprior}
}
\end{figure}

We modelled the distribution in flux and redshift space of the background source population. We started from the observed galaxy distribution in rest-frame UV ($1500\AA$) luminosity and redshift space, as measured by \citet{Par++16}. We discuss the impact of the choice of template in \Sref{sc:Disc}. Then, for a given value of the UV luminosity and the redshift, we computed the \IE flux using the spectral energy distribution template \texttt{SB1}, from the Kinney--Calzetti atlas, as provided by \texttt{LePhare}. 
Figure~\ref{fig:sourceprior} (filled contours) shows the resulting prior distribution in redshift and \IE-flux space, $\prsource(\zsource,\sourceflux)$.

\begin{figure}
\includegraphics[width=\columnwidth]{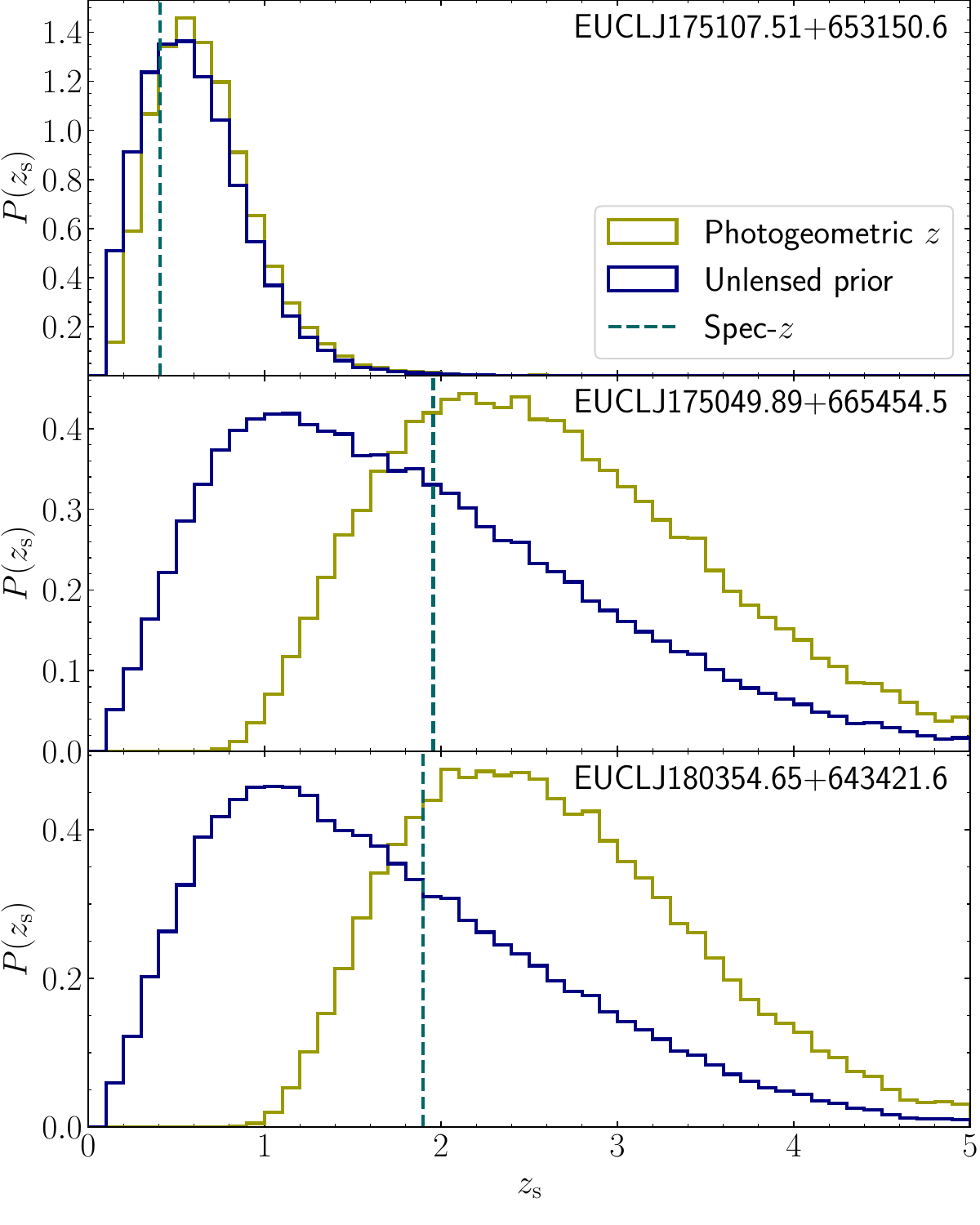}
\caption{
Photogeometric redshifts of the three lensed sources with a spectroscopic redshift.
The blue colour histogram shows the prior probability $\prsource(\zsource|\sourceflux)$.
The gold colour histogram shows $P(\zsource|\zlens,\tein,\sourceflux,\fluxset,\teff)$, obtained from \Eref{eq:photogeoz} by marginalising over the parameters $\hyperpars$, given the posterior probability of the model.
\label{fig:photogeoz}
}
\end{figure}

Given $\prsource(\zsource,\sourceflux)$, the posterior probability of a lensed source given a model population $\hyperpars$ and the other observables is 
\begin{align}
& P(\zsource| \tein, \sourceflux,\fluxset,\teff,\hyperpars) \propto \prsource(\zsource|\sourceflux) \nonumber \\
& \times \int P(\tein|\psilens,\zsource)\, P(\fluxset,\teff|\psilens)\, \prlens(\psilens|\hyperpars)\, \crosssect(\psilens,\psisource)\, \diff\psilens \label{eq:photogeoz}\,,
\end{align}
which we refer to as the photogeometric redshift of the source.
Figure~\ref{fig:photogeoz} shows the photogeometric redshift of the three lensed sources with spectroscopic redshifts, given the redshift of the lens, and marginalised over the uncertainty on the model parameters $\hyperpars$.
In all cases, the photogeometric redshift is consistent with the spectroscopic redshift.
Figure~\ref{fig:photogeoz} also shows the prior on $\zsource$ conditional on the flux, $\prsource(\zsource|\sourceflux)$.
In the case of EUCL\,J175107.51+653150.6, this term dominates the posterior. This is because that source is particularly bright, and flux information alone rules out solutions with a high redshift.
For the other two sources, the addition of the Einstein radius constraint shifts the distribution to higher redshifts, because lower redshift solutions require a particularly large, hence unlikely, lens mass. 

\subsection{Lensing cross-section}\label{ssec:crosssect}

In order to compute the normalisation of the probability distribution, \Eref{eq:norm2}, we need an expression for the lensing cross-section.
The lensing cross-section depends primarily on the mass distribution of the lens, on its redshift, on the redshift of the background source, and on the flux of the source. However, it also depends on the surface brightness distribution of the foreground galaxy, because our definition of a golden lens requires a minimum contrast between the source and lens light.

We can compute $\crosssect$ by means of simulation. For a given set of properties of the lens galaxy and of the background source, we can draw a large number of source positions, sampled uniformly in a sufficiently large region of the sky, and, for each configuration, create a mock \Euclid image. We can then check, for each image, if the resulting lens satisfies our golden lens definition. The final value of $\crosssect$ is the ratio between the number of golden lenses to the total number of generated images, multiplied by the source plane area from which the source positions were sampled. However, doing so for every possible combination of lens and source parameters, and for any possible value of the model parameters $\hyperpars$, is unpractical. Therefore, we made a few simplifying assumptions. 
We fixed the axis ratio of the lens light and lens mass to the same value of $q=0.8$ \citep[the lensing cross-section has a very mild dependence on axis ratio, as shown by][]{Son++23}. Then, for the purposes of generating images of background sources, we assumed a fixed scaling relation between the angular size of the source and its flux,
\begin{equation}
\logten{\left(\theta_{\mathrm{s,e}}/1\arcsecond\right)} = -0.850 + 0.418\logten{\left(\sourceflux/\mujansky\right)}\,.
\end{equation}
We obtained the above relation by fitting the flux--size distribution of the background sources in our sample with a linear model. Finally, we made the assumption that the lensing cross-section scales with Einstein radius as
\begin{equation}\label{eq:teinscaling}
\frac{\crosssect(\tein,\lensflux,\teff,\sourceflux)}{\crosssect(\teinref,\lensflux,\teff,\sourceflux)} = \left(\frac{\tein}{\teinref}\right)^2\,,
\end{equation}
where $\teinref$ is a reference value of the Einstein radius.
Equation (\ref{eq:teinscaling}) says that, for two lenses with identical lens light distribution and source flux, the ratio of their lensing cross-section is equal to the ratio of their Einstein radius squared. This is the theoretical scaling of the lensing cross-section for an SIE lensing a point source, and we expect it to hold in our regime as well. The investigation of the lensing cross-section of lenses with extended lensed sources by \citet{Son++23} supports this claim.
One important exception to \Eref{eq:teinscaling} is if $\tein$ falls below our minimum Einstein radius of $\ang{;;0.7}$, in which case $\crosssect=0$.

\subsection{Practical implementation}\label{ssec:practical}

We sampled the posterior probability distribution $P(\hyperpars|\data)$ with a Markov chain Monte Carlo (MCMC). At each draw of $\hyperpars$ and for each lens, this requires computing integrals of the kind of \Eref{eq:integrals}. To simplify calculations, we assumed that all of the observables, including the Einstein radius, are measured exactly. This is a good approximation, as the typical uncertainty on $\tein$ is on the order of a few $\%$ \citep{Eth++22}, which is much smaller than the uncertainty on the mass model parameters, and also much smaller than the intrinsic scatter in $\mfive$.
The contribution to the likelihood of a lens without a spectroscopic redshift of neither the lens nor the source is
\begin{align}
P(\tein,\fluxset,\teff,\sourceflux|\hyperpars) = & \int \prlens[\zlens,\fluxset,\teff,\mfive(\tein,\zlens,\zsource)|\hyperpars] \nonumber \\
& \times\prsource(\zsource|\sourceflux)\, \left|\frac{\diff\mfive}{\diff\tein}\right|\label{eq:integral2}\diff\zlens \diff\zsource\,,
\end{align}
where we used $\psel=1$ (i.e. we assumed the lens sample to be effectively complete),
and the last factor in the integrand is the result of the integration over $\mfive$, which we performed analytically in virtue of the likelihood in the Einstein radius being a Dirac delta function.
We computed the integrals above (one per each lens) by Monte Carlo integration, drawing samples of $\zlens$ from the lens photo-$z$ posterior probability, and samples of $\zsource$ from the distribution in source redshift, conditional on the source flux. For each $\zlens<\zsource$ pair, we computed the value of $\mfive$ necessary to match the observed Einstein radius. For $\zlens>\zsource$, the integrand is zero.

To compute the normalisation of the probability distribution, \Eref{eq:norm2}, we proceeded as follows. 
We drew a sample of $10\,000$ lens and source pairs from $\prsl(\psilens,\psisource|\hyperparsref)$, where $\hyperparsref$ is a reference value of the model parameters. For each lens-source pair, we computed the lensing cross-section following the procedure described in \Sref{ssec:crosssect}. Then, for any given value of $\hyperpars$, we drew samples of $(\psilens,\psisource)$ from the same random seed. This created a new sample with identical lens light and source properties, and with a shifted distribution in $\mfive$. We made use of the $\crosssect$ scaling approximation of \Eref{eq:teinscaling} to compute the lensing cross-section for this new sample. Finally, we computed the integral of \Eref{eq:norm2} by Monte Carlo integration. We iterated this procedure, each time setting $\hyperparsref$ to a value close to the peak of the posterior probability. The final results converged after two iterations.

We assumed uniform priors on all of the mass model parameters, over the ranges listed in \Tref{tab:inference}.
We fixed the parameters of the photo-$z$ model and of the distribution in redshift, luminosity and size to the median of their marginal posterior probability. 

\section{\label{sc:Results} Results}

\begin{table*}
\caption{Inference on the mass model parameters. }
\smallskip
\label{tab:inference}
\smallskip
\centering
\begin{tabular}{ccccp{9cm}}
Parameter & Prior & Posterior ($68\%$) & Eq. def. & Description \\
\hline
\hline
$\mu_{M,0}$ & $U(10.5,11.8)$ & $11.300 \pm 0.021$ & (\ref{eq:mumfive}) & Average $\logten{(M_5\,h\,M_\odot^{-1})}$ for a galaxy at $z=0.5$, with $L_r=10^{10.7}L_{r,\odot}\,h^{-2}$ and $\logten{(\reff\,h/\mathrm{kpc})}=0.5$.\\
$\sigma_M$ & $U(0.02,0.2)$ & $0.099 \pm 0.010$ & (\ref{eq:mfive}) & Intrinsic scatter in $\logten{(M_5\,h\,M_\odot^{-1})}$ at fixed redshift, luminosity, and size.\\
$\beta_L$ & $U(-3,3)$ & $0.71 \pm 0.07$ & (\ref{eq:mumfive}) & Power-law scaling of $\mfive$ with luminosity.\\
$\beta_R$ & $U(-3,3)$ & $-0.15 \pm 0.07$ & (\ref{eq:mumfive}) & Power-law scaling of $\mfive$ with size.\\
$\beta_z$ & $U(-3,3)$ & $-0.33 \pm 0.05$ & (\ref{eq:mumfive}) & Power-law scaling of $\mfive$ with redshift.\\
\end{tabular}
\tablefoot{The second column reports the assumed prior probability distribution, where the notation $U(x,y)$ indicates a uniform distribution between $x$ and $y$. The third column reports the median and $68\%$ credible interval of the marginal posterior probability. The fourth column indicates the equation where each parameter is first defined.}
\end{table*}

\begin{figure*}[htbp!]
\includegraphics[width=\textwidth]{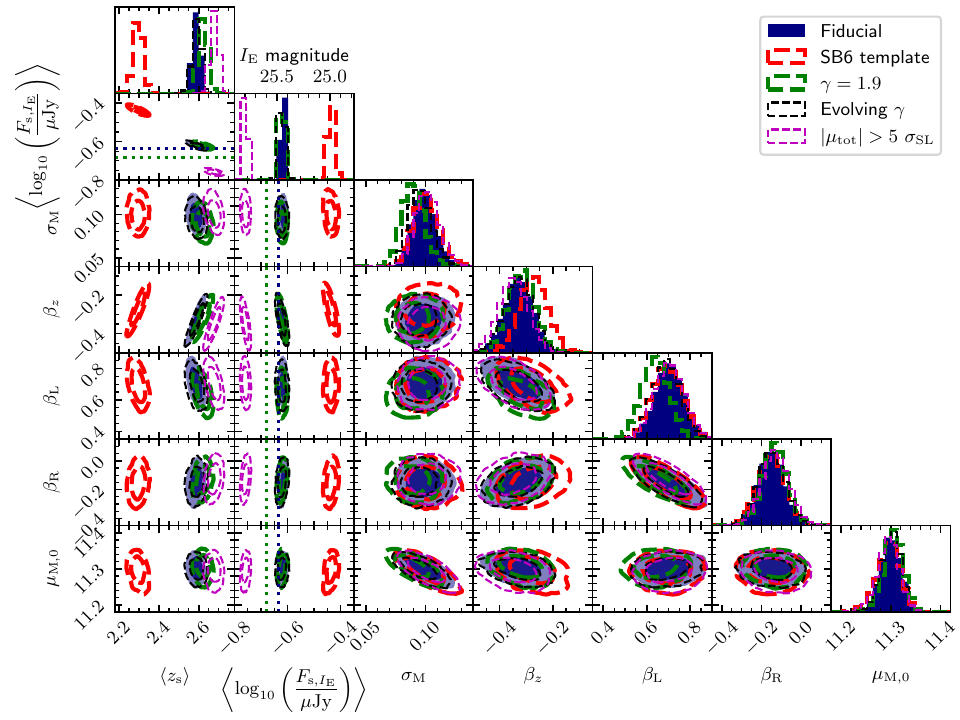}
\caption{Posterior probability distribution of the parameters of the model. The first two columns show the posterior probability of the average redshift and flux of the lensed sources (predicted from the model, and not fitted for). Blue filled contours show the fiducial inference, obtained under the SIE model assumption. Green dashed contours are obtained with a power-law density profile with $\gamma=1.9$. Black dashed contours are obtained with a power-law density profile with a slope evolving as $\gamma(z)=2 - 0.3(z-0.5)$. Red dashed contours are obtained by repeating the fiducial analysis, but with a different choice for the spectral template used to describe the population of background sources. These three alternative models are introduced in \Sref{sc:Disc}. 
Contour levels correspond to 68\% and 95\% enclosed probability.
Dotted lines indicate the observed average source flux, under the assumption of an SIE (blue colour) or $\gamma=1.9$ lens model (green colour).
}
\label{fig:cp}
\end{figure*}

Table~\ref{tab:inference} reports the median and $68\%$ credible interval of the marginal posterior probability distribution of the model parameters.
The full probability distribution is shown in \Fref{fig:cp} by means of pairwise projections.
The parameter $\mu_{\mathrm{M},0}$, which describes the average $\logten{(\mfive\,h/M_\odot)}$ at a reference luminosity, size, and redshift, is recovered with a $68\%$ uncertainty of $0.02$. This implies a $5\%$ uncertainty on the mass-to-light ratio. 
We see a negative covariance between $\mu_{\mathrm{M},0}$ and the intrinsic scatter parameter $\sigma_\mathrm{M}$, which can be explained as follows. 
While our model describes the general population of galaxies, lensing data samples only the high mass tail of the galaxy distribution.
A model with an overall lower mass and higher intrinsic scatter can produce a lens population with similar Einstein radius distribution to that of a model with a higher mass and smaller scatter. Therefore, if the only lensing-related constraints are those provided by $\tein$, these two scenarios are degenerate.

We detected a positive correlation of $\mfive$ with luminosity, $\beta_\mathrm{L}=0.71\pm0.07$. 
This value is consistent with the scaling between stellar mass and the total projected mass enclosed within $3.5\,h^{-1}\,\mathrm{kpc}$ measured by \citet{Son24}.
We found a preference for a negative value of the mass--size correlation parameter, $\beta_\mathrm{R}=-0.15\pm0.07$, although it is strongly degenerate with the luminosity dependence parameter, $\beta_\mathrm{L}$.

\begin{figure}
\includegraphics[width=\columnwidth]{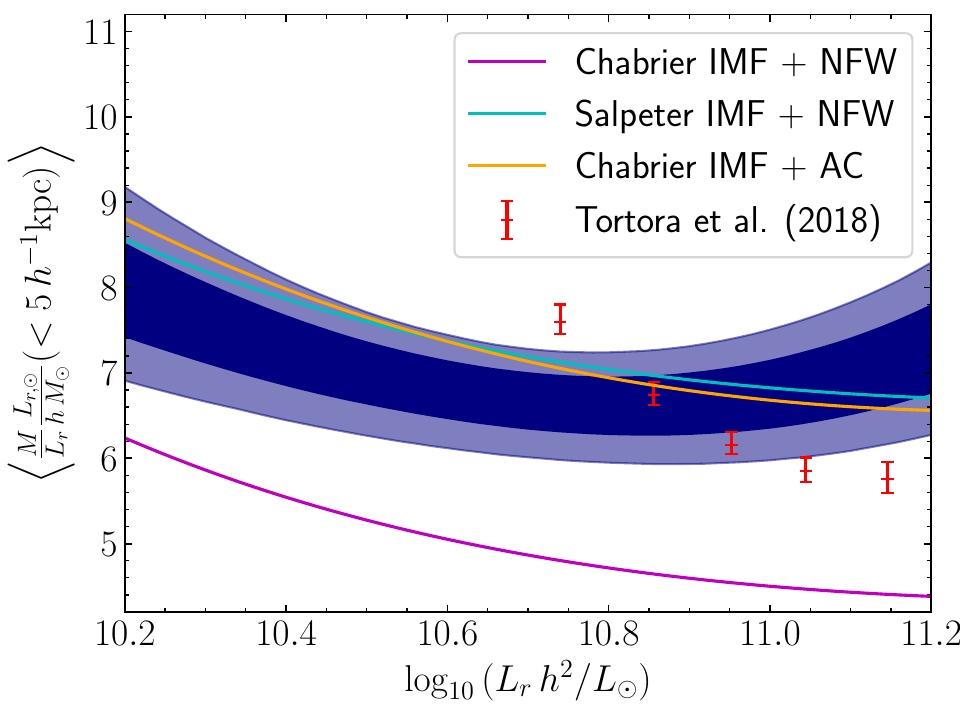}
\caption{
Mean mass-to-light ratio, integrated within an aperture of $\fivekpc$, as a function of galaxy luminosity (blue bands). The bands enclose $68\%$ and $95\%$ of the probability. The model assumes a fixed redshift $\zlens=0.5$, and average size for the luminosity.
The three curves show theoretical predictions from models consisting of stars and a dark matter halo. We described the stars with an SPS model, obtained from \texttt{Galaxev} \citep{B+C03}. We assumed solar metallicity, an exponentially declining star formation history with timescale $\tau=1\,\mathrm{Gyr}$, $7\,\mathrm{Gyr}$ since the initial burst, and minimum dust ($\tau_V=0.04$). We adopted two choices for the stellar IMF, as indicated in the legend. For the dark matter halo, we assumed an NFW profile or an adiabatically contracted (AC) profile.
\label{fig:ml_vs_lum}
}
\end{figure}

We combined these results to compute the mass-to-light ratio of elliptical galaxies as a function of luminosity. In particular, we computed the mean ratio between mass and luminosity, each within an aperture of $\fivekpc$, of galaxies at $\zlens=0.5$, and with average size for their luminosity. The result is shown in \Fref{fig:ml_vs_lum}.
At a luminosity $\logten{(L_r\, h^2/L_\odot)}=10.7$ (close to the average of the golden lens sample), the mean mass-to-light ratio within $\fivekpc$ is $(6.7\pm0.3)\,M_\odot\, h\,L_{r,\odot}^{-1}$.
For small luminosities, the mass-to-light ratio is a decreasing function of $L_r$. This is a consequence of the luminosity-dependence parameter $\beta_\mathrm{L}$ being smaller than one. 
In galaxies of decreasing luminosity, an aperture of $\fivekpc$ corresponds to a relatively larger radius compared to the half-light radius, and therefore to a region more and more dominated by dark matter. 
The upturn in $M/L_r$ at high $L_r$ is instead driven by the steepening of the luminosity-size relation. At a given luminosity, a larger half-light radius means that the fraction of light enclosed within a fixed physical aperture decreases, while $M/L_r$ increases.

In order to interpret these results, we show in \Fref{fig:ml_vs_lum} theoretically predicted mass-to-light ratios, obtained from models consisting of the sum of a stellar component and a dark matter halo. We modelled the stellar component with stellar population synthesis (SPS) models obtained from the software \texttt{Galaxev} \citep{B+C03}. We assumed solar metallicity, an exponentially declining star formation history, and minimal dust. We adopted both a Chabrier and a Salpeter stellar initial mass function (IMF). For each value of the luminosity, we obtained the corresponding stellar mass given the SPS model. 
Then, using the stellar-to-halo mass relation measured by \citet{Shu++22}, we set the dark matter halo mass $M_h$ to the average $\logten{(M_h/M_\odot)}$ of galaxies of the given stellar mass. 
We considered dark matter halos with a Navarro, Frenk \& White \citep[NFW;][]{NFW97} profile with a mass-concentration relation from \citet{D+M14}, as well as adiabatically contracted profiles, obtained with the prescription of \citet{Blu86}.
The models straddle the range of values of mass-to-light ratio allowed by the data. This indicates that our measurements can potentially rule out regions in the model parameter space defined by the stellar IMF and the dark matter profile. However, given that we set the stellar population parameters arbitrarily when building these models, we refrain from drawing quantitative conclusions from this comparison.

Figure~\ref{fig:ml_vs_lum} also shows dynamical measurements obtained from the work of \citet{Tor++18}, for comparison. \citet{Tor++18} estimated the mass of a large sample of massive $[\log_{10}{(M_*\,h^2/M_\odot^{-1})}>10.95]$ early-type galaxies, using their stellar velocity dispersion as a constraint. They, too, assumed an isothermal density profile, and used the spherical Jeans equation to constrain their mass. The \citet{Tor++18} values are, on average, consistent with our measurements. However, they show a significantly steeper decreasing trend with luminosity. The reason for the sharper decrease reflects a stronger sensitivity of the dynamical estimates to the half-light radius. 
For a singular isothermal sphere, $\mfive\propto \sigma_{\mathrm{SIS}}^2$, where $\sigma_{\mathrm{SIS}}$ describes the velocity dispersion of the massive particles that form the distribution. In the context of Jeans modelling, $\sigma_{\mathrm{SIS}}$ is, to first approximation, similar to the velocity dispersion of the tracer particles (i.e. the stars). At fixed stellar mass, the stellar velocity dispersion is a decreasing function of $\reff$ \citep[e.g.][]{Ber++20}. As a result, the inferred $\mfive$ appears to decrease with increasing $\reff$. By contrast, our measurement shows a very mild dependence of $\mfive$ on $\reff$, quantified by the parameter $\beta_\mathrm{R}=-0.15\pm0.07$. Since higher-luminosity galaxies have on average larger sizes, this different sensitivity to $\reff$ is responsible for the difference in slope observed in \Fref{fig:ml_vs_lum}.
A problem with the dynamical measurements is that, in practice, the stellar velocity dispersion depends not only on the overall mass, but also on the slope of the density profile. At fixed $\mfive$, it is larger for galaxies with a steeper density slope. 
More compact galaxies have a steeper density slope \citep{Son++13b}, and therefore a larger stellar velocity dispersion. But, under the isothermal assumption, the dependence on the density slope is absorbed into a dependence on $\reff$, introducing a bias. In contrast, our lensing-based measurements are less sensitive to the choice of the density profile, as we show in \Sref{sc:Disc}. Nevertheless, the fact that the dynamical data points are, on average, in agreement with our estimate, is an important consistency check.

\begin{figure}
\includegraphics[width=\columnwidth]{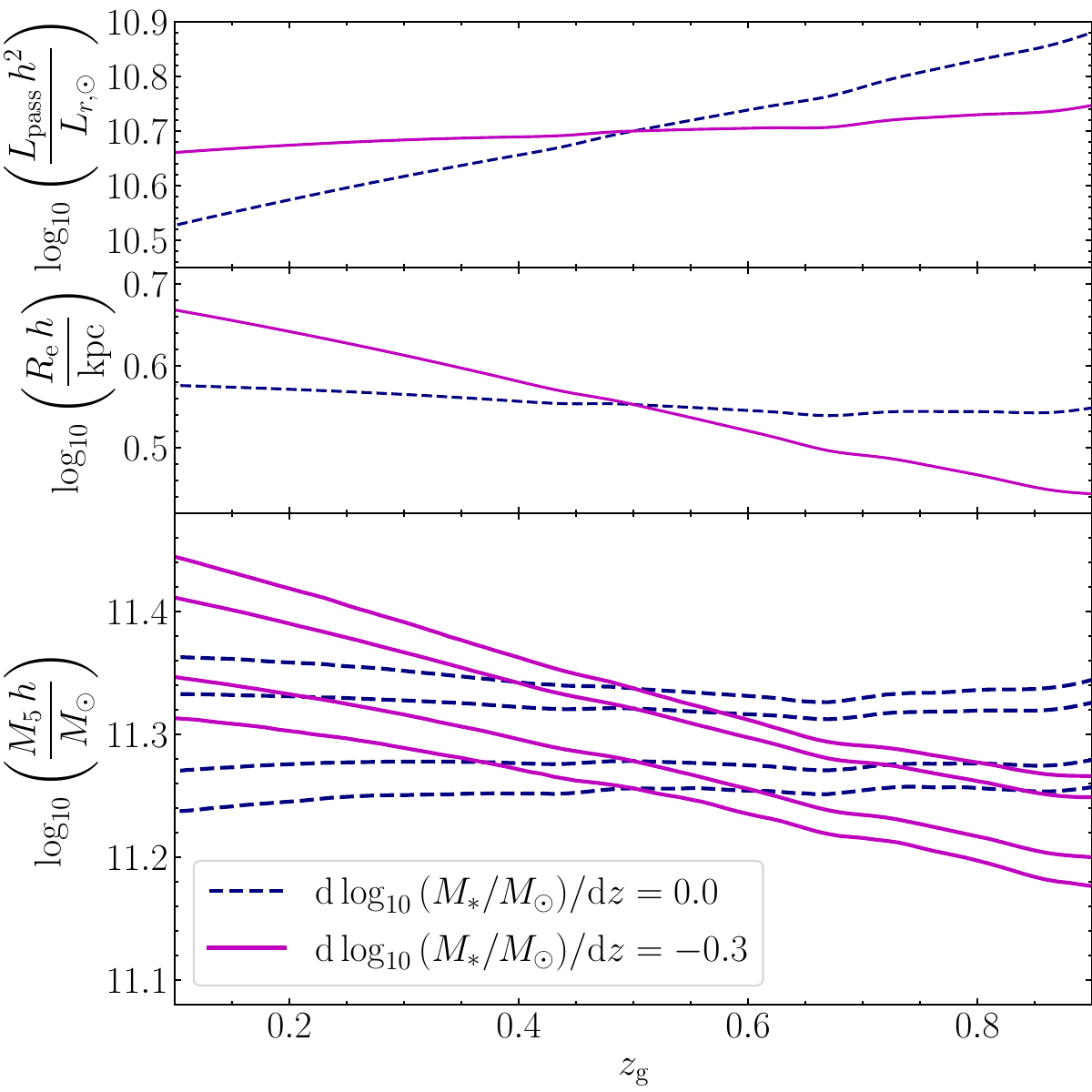}
\caption{
Evolution of individual galaxies, under the assumption of a purely merger-induced growth. Two stellar mass growth rates are considered.
\emph{Top}: luminosity evolution, assuming a simple stellar population formed at $z=2.5$, with solar metallicity. We adopted $H_0=70\,\mathrm{km}\,\mathrm{s}^{-1}\,\mathrm{Mpc}^{-1}$ to convert stellar ages to redshift.
\emph{Middle}: average size of galaxies with luminosity equal to $L_{\mathrm{pass}}$.
\emph{Bottom}: average $\logten{(\mfive\, h/M_\odot)}$ of galaxies with luminosity equal to $L_{\mathrm{pass}}$, and average size for the luminosity, as a function of redshift. Regions correspond to 95\% and 68\% enclosed probability.
\label{fig:passevol}
}
\end{figure}

We also found a negative correlation between redshift and mass, at fixed luminosity. This is expected, because the luminosity of elliptical galaxies decreases with time, as their stellar populations become older. The result is that, at the same luminosity, lower redshift galaxies are more massive.
Rather than comparing the properties at fixed luminosity, a more interesting description is one that follows the evolution of $\mfive$ along the evolutionary tracks of galaxies. 
However, matching progenitors and descendants in a sample of galaxies is not trivial. 
Elliptical galaxies grow mainly as a result of mergers, but the amount of growth is uncertain \citep{Muz++13,Bun++17,Ada++21}.
We considered two extreme scenarios. One in which there is no growth between $z=1$ and $z=0$, and one in which the stellar mass grows as $\diff\logten{(M_*/M_\odot)}/\diff z = -0.3$ (roughly a factor of two increase over the same period).
To further simplify the problem, we assumed that the stellar population of the elliptical galaxies and of the accreted stars were all generated in a single burst at $z=2.5$.
We used \texttt{Galaxev} to compute the luminosity evolution in the two cases.
The result is shown in \Fref{fig:passevol}. The top panel shows $L_{\mathrm{pass}}(z)$, which we normalised so that $\logten{[L_{\mathrm{pass}}(0.5)\,h^2/L_{r,\odot}]}=10.7$. In the no-growth scenario, the luminosity drops by $0.36$~dex from $z=1$ to $z=0$, while in the rapidly growing case it decreases by $0.10$~dex. 
The bottom panel shows the corresponding evolution in the average $\mfive$, assuming that, at each redshift, the size stays equal to the average for galaxies of the same luminosity (middle panel).
In the case of no growth, the data are consistent with a non-evolving mass. The inferred variation in the average $\mfive$ between $z=1$ and $z=0$ is $(-0.01\pm0.05)\,\mathrm{dex}$. In the rapidly growing case, the variation is $(0.18\pm0.05)\,\mathrm{dex}$. 
Although we cannot distinguish between these two scenarios, we can conclude that, if there is any merger-induced growth between $z=1$ and $z=0$, $\mfive$ increases along the progenitors. Using the rapidly growing case as a reference, we estimate the response in $\mfive$ to an increase in stellar mass to be approximately $\diff\logten{(\mfive/M_\odot)}/\diff\logten{(M_*/M_\odot)} = 0.54\pm0.15$\footnote{Adopting a different stellar mass growth rate does not change this result significantly.}.
This measurement can be used to test models of the merger evolution of elliptical galaxies. For example, \citet{Liu++25} showed that major and minor mergers have a different impact on the inner stellar mass distribution of elliptical galaxies, for the same amount of overall stellar mass growth.

\begin{figure}
\includegraphics[width=\columnwidth]{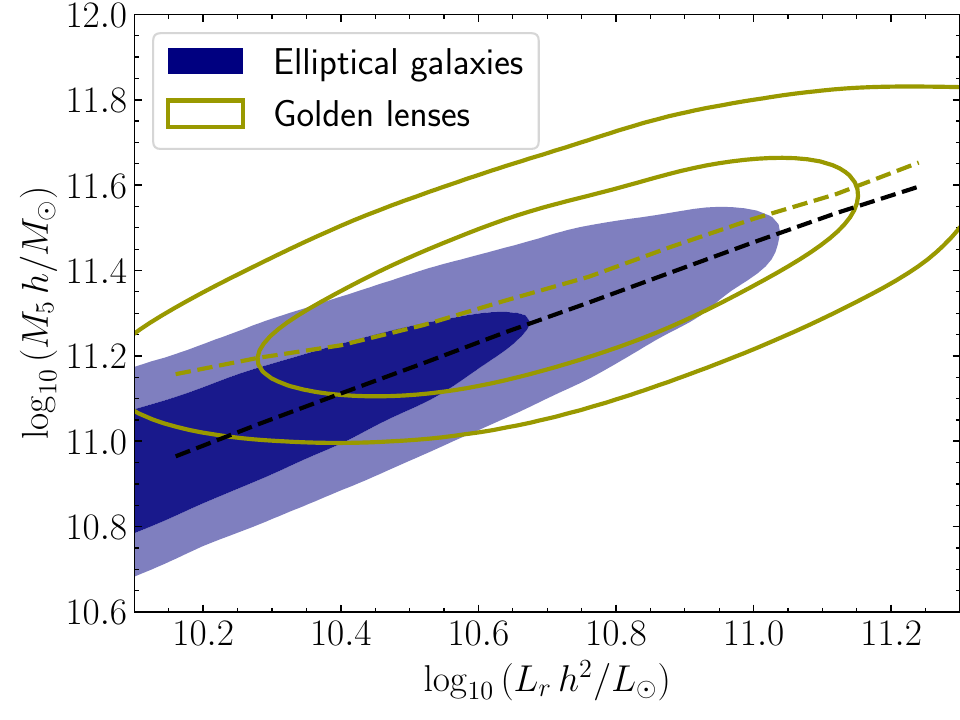}
\caption{
Distribution in luminosity and $\mfive$ of elliptical galaxies and golden lenses. 
For each sample, contours show the distribution predicted for the best-fit values of the model parameters (i.e. the median values listed in \Tref{tab:inference}), for galaxies at $\zlens=0.5$.
Dashed lines indicate the mean of each distribution.
\label{fig:lum_vs_m5}
}
\end{figure}

Finally, we investigated how the lens population differs from the general population of galaxies in terms of mass. 
The former is given by the left-hand side of \Eref{eq:one}, while the latter is described by $\prlens(\psilens)$.
Figure~\ref{fig:lum_vs_m5} shows the predicted distribution of $\mfive$ as a function of luminosity, at a reference redshift $\zlens=0.5$. At fixed luminosity, lenses are on average more massive than regular galaxies (dashed lines in \Fref{fig:lum_vs_m5}). The difference is $0.08$~dex at $\logten{(L_r\,h^2/L_{r,\odot})}=10.6$, and $0.05$~dex at the high-luminosity end. Figure~\ref{fig:lum_vs_m5} only shows the distribution obtained from the best-fit model parameter values. In reality, the amplitude of the difference between the two distributions depends on the value of the intrinsic scatter parameter (larger for larger $\sigma_M$). Nevertheless, this result indicates that particular care must be taken in interpreting strong lensing measurements directly, especially if a precision of $20\%$ or better is desired.

All results in this paper are based on galaxies with no neighbours within $3\arcsecond$ in projection. Therefore, they should be interpreted as applying to the subset of elliptical galaxies with no close projected companions.

We compared our results with the lensing-based measurements of \citet{Son24}, who obtained the total projected mass of early-type galaxies within $3.5\,h^{-1}\,\mathrm{kpc}$, as a function of stellar mass, using strong lenses from the Sloan ACS Lens Survey \citep[SLACS;][]{Bol++06}.
To do so, we first extrapolated the \citet{Son24} measurement to an aperture of $\fivekpc$, using the power-law density profile model of \citet{Son24}. Second, we obtained from \citet{Aug++09} the $V$-band luminosity corresponding to the stellar mass of each SLACS lens. Third, we converted from $V$- to $r$-band luminosity using a spectral template with colour $\logten{(L_r/L_g)}=0.35$, which is the typical value for elliptical galaxies at $z=0.2$ (\Fref{fig:physical}). The result is that the average $\logten{(\mfive\, h/M_\odot)}$ at $\logten{(L_r\, h^2/L_{r,\odot})}=10.67$ is $11.33\pm0.01$. Our model predicts a value of $11.38\pm0.03$ for galaxies of the same luminosity and redshift, therefore is within two standard deviations of the \citet{Son24} value.

\begin{figure}
\includegraphics[width=\columnwidth]{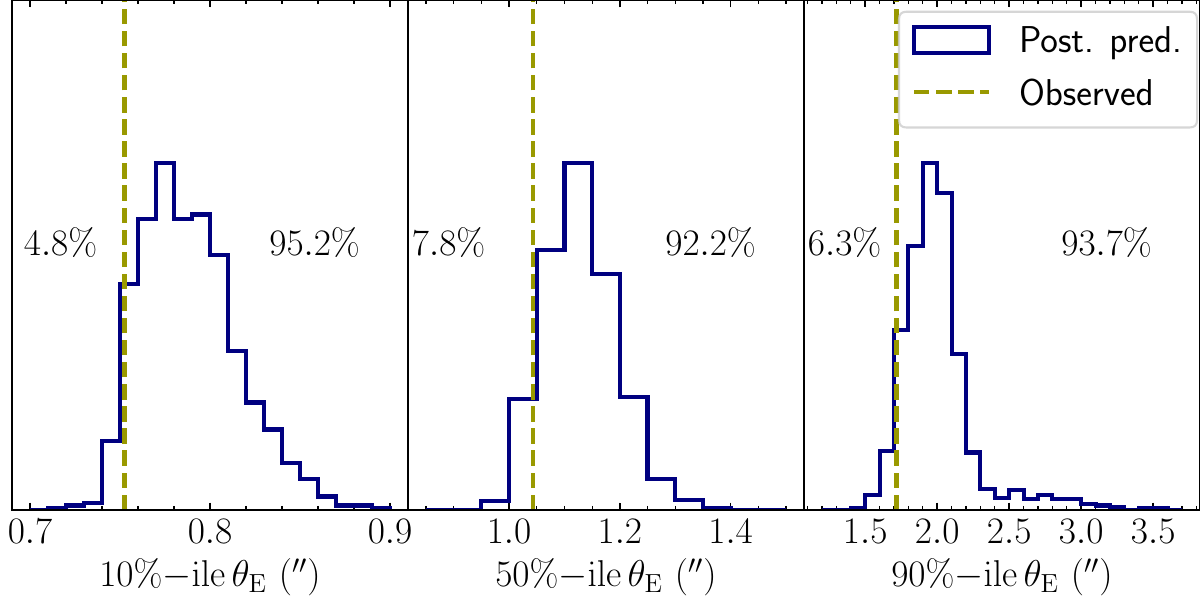}
\caption{
Goodness-of-fit posterior predictive test.
Each panel shows a posterior predicted percentile of the Einstein radius distribution, as indicated on the horizontal axis. The percentages indicate the fraction of samples for which the posterior predicted value is smaller or larger than the observed value.
\label{fig:pptest}
}
\end{figure}

We checked the goodness of fit of our model by investigating how well it can reproduce the observed distribution in Einstein radius, by means of a posterior predictive test. In particular, we used the $10\%$-, the $50\%$-, and the $90\%$-ile of the distribution in $\tein$ as test quantities.
We drew samples from the posterior probability of the model parameters. For each draw, we generated a population of $\Ngold$ golden lenses and measured the three test quantities.
Figure~\ref{fig:pptest} shows the resulting posterior predicted distributions.
The percentages on the left and right of each panel indicate the probability that a smaller or larger value than the observed one would have occurred by chance, under the assumption of the model being correct.
Although the model tends to predict larger values of the percentiles of the Einstein radius distribution, the chance of observing even smaller values by chance is reasonably large ($5.7\%$ for the 90\%-ile, which is the test quantity with the worst agreement). Therefore the test gives us no reason to question the goodness of fit of the model to the Einstein radius distribution.

\section{\label{sc:Disc} Discussion}

Our study is based on a sample of lenses with minimal spectroscopic coverage, which limited our ability to obtain precise measurements of mass on an individual lens basis. We overcame this problem with statistics, and with two key modelling choices.
The first one is the assumption of a singular isothermal density profile for the mass distribution. To investigate the impact of this assumption, we repeated the analysis with a power-law density profile, $\rho(r)\propto r^{-\gamma}$, with $\gamma$ set to plausible values that are consistent with measurements from the literature.
One caveat is that our inference of the fluxes of the lensed sources also depends on the choice of the density profile, because it relies on a model for the lensing magnification.
Our measurement of $\sourceflux$ was obtained using an SIE model.
In order to be self-consistent with the new power-law assumption, we multiplied them by $\gamma-1$, which is the ratio between the radial magnification of an SIE and that of a power-law lens, evaluated at the Einstein radius.
In other words, we accounted for the fact that, if we had modelled our lenses with a power-law model, we would have inferred smaller intrinsic source fluxes, because of its larger radial magnification compared to an SIE.

We first fixed $\gamma=1.9$ for all galaxies. This is close to the value inferred by \citet{Bol++12} for galaxies at $z=0.6$.
The resulting posterior probability distribution is shown in \Fref{fig:cp} as dashed green curves.
The inference on the average mass parameter, $\mu_{M,0}$, is virtually unchanged, while that on the other parameters is consistent with the fiducial model. 
We also considered a power-law model with a density slope that evolves as $\gamma(z) = 2.0 - 0.3(z-0.5)$, which is very close to the trend measured by \citet{Son++13b} over the range $0<z<0.8$. As \Fref{fig:cp} shows, the inferred parameter values are still unchanged.
Neither of these results are fortuitous coincidences, but rather the consequence of having chosen $\fivekpc$ as the reference scale on which to measure the total mass. Because the median Einstein radius of the golden lenses is very close to $\fivekpc$, the data constrain $\mfive$ directly, with a very small sensitivity to the choice of the density profile.
More extreme deviations from an isothermal profile are in tension with existing constraints from lensing and stellar dynamics \citep{Son++13b}.
We conclude that the isothermal assumption is currently not limiting the accuracy of our measurement.

We could also modify our density profile with a mass-sheet transformation, 
\begin{equation}
\Sigma(R) \rightarrow \lambda_{\mathrm{mst}}\Sigma(R) + (1-\lambda_{\mathrm{mst}})\Sigma_{\mathrm{cr}}\,,
\end{equation}
which leaves our lensing observables unchanged \citep{S+S13}.
For small projected distances from the Einstein radius, the projected mass profile of a mass-sheet-transformed isothermal model can be approximated by a power law. Therefore, we expect the inference to be robust also with respect to modest mass-sheet transformations ($\lambda_{\mathrm{mst}}$ close to $1$).

One additional caveat with the isothermal assumption, and with the tests described so far, is that we did not allow for scatter in the density profile. Given that the Einstein radii of our lenses have a non-zero spread around $\fivekpc$, a scatter in the density slope $\gamma$ would be absorbed by the parameter that describes the intrinsic scatter in $\mfive$, $\sigma_\mathrm{M}$. 
Therefore, we urge caution in taking our measured value of $\sigma_\mathrm{M}$ at face value.
For a quantitative estimate of the degeneracy between the scatter in density slope and mass, we refer the reader to the analysis of the SLACS sample of \citet[][figure 4]{Son24}. 

The second critical assumption of our model is the choice of the prior in flux--redshift space of the population of background sources, which we based on a measurement of the UV luminosity function of galaxies from \citet{Par++16}. 
Our implicit assumption is that the method that we used to measure source fluxes is consistent with that used by \citet{Par++16}. This assumption, however, is difficult to verify and could introduce biases in the inferred redshift distribution of the background sources.
Moreover, we assumed a fixed spectral template, \texttt{SB1}, to convert from rest-frame UV to observed \IE flux. 
To assess the sensitivity of our results to this particular choice, we repeated the analysis while using the template \texttt{SB6}, which has a significantly redder spectrum. The resulting posterior probability is shown in \Fref{fig:cp} (red dashed contours). The inferred average lensed source redshift is significantly lower than that of the fiducial analysis. However, the values of the model parameters show very small differences, indicating that the choice of source spectral template is not a limiting factor in our analysis. This is not surprising, given that the critical surface mass density is a slowly varying function of source redshift. 
The \texttt{SB1} template produces good agreement between the posterior predicted average source flux and the observed one (\Fref{fig:sourceprior}). 
By contrast, when using the \texttt{SB6} template, the posterior predicted average source flux is larger than the observed one by $0.14$~dex (\Fref{fig:cp}). For this reason, we chose the \texttt{SB1} template for our fiducial analysis. Nevertheless, our understanding of the background source population is still subject to systematic uncertainties.
We have an ongoing spectroscopic follow-up programme. With the aid of more spectroscopic observations, we plan to measure the distribution of the background source population directly on \Euclid data.

\begin{figure}
\includegraphics[width=\columnwidth]{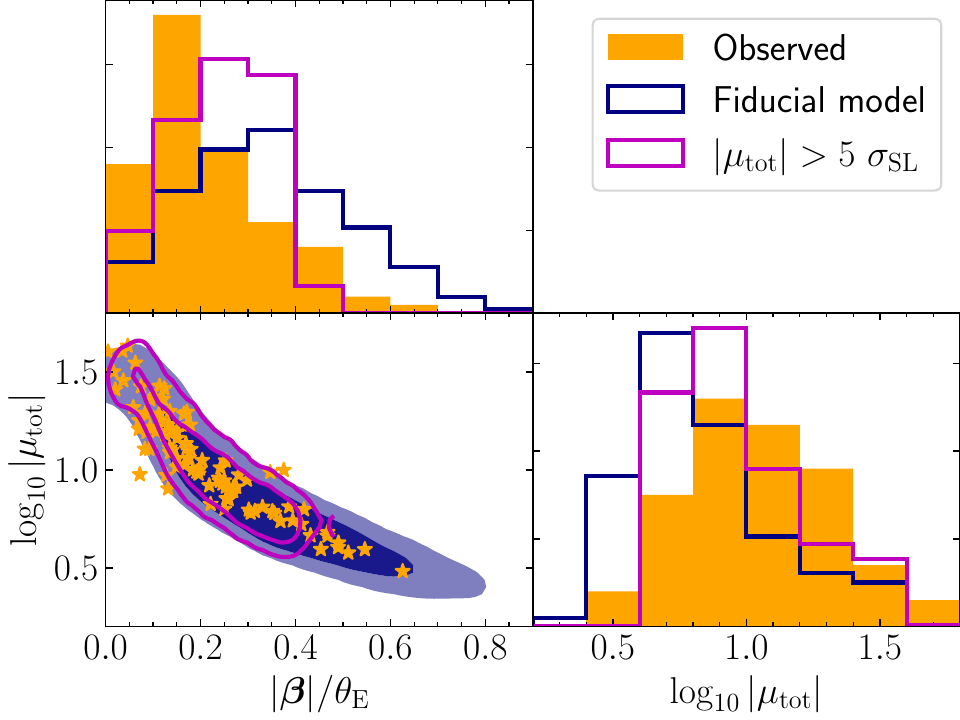}
\caption{
Distribution in total magnification and projected distance between the source and the lens, normalised by the Einstein radius.
Orange stars and histograms: observations. Blue: prediction from the fiducial model, with each population parameter fixed to the median value of its marginal posterior probability. Magenta: prediction from a model with a modified lensing cross-section, with $\crosssect=0$ for lenses with total magnification $|\mu_{\mathrm{tot}}|<5$.
\label{fig:betamu}
}
\end{figure}

Another assumption that we made is that of the effective completeness of our sample of golden lenses. 
That is, we assumed that our sample is missing golden lenses at random, with no correlation between the properties of a lens and the probability of us including it in our sample. One way to test this assumption is by examining additional aspects of the properties of the lenses, such as the inferred source position $\sourcepos$ and the total magnification $|\mu_{\mathrm{tot}}|$. We defined the latter as the ratio between the total flux of the lensed source and the inferred intrinsic flux. Figure \ref{fig:betamu} shows $|\mu_{\mathrm{tot}}|$ as a function of $|\sourcepos|/\tein$ for the observed sample, along with the distribution predicted with our fiducial model (blue contours). 
Lenses with a larger value of $|\sourcepos|/\tein$ have a more asymmetric image configuration, with a main arc well outside of the critical curve and a fainter counter-image near the lens centre, and smaller total magnification. At the oposite limit, lenses with $|\sourcepos|/\tein\approx0$ produce Einstein rings, with a large magnification.
Qualitatively, the model reproduces this trend. However, it predicts a tail of low-magnification and large source-lens separation that is not observed in our sample (most clearly seen in the histograms of \Fref{fig:betamu}).
Such lenses must exist in reality, but are evidently identified with a lower efficiency compared to the highly-magnified ones. 
This means that our assumption of the golden lens sample being effectively complete is not entirely accurate. In order to understand the impact of this inaccuracy on our measurements, we repeated our analysis with an alternative model for the strong lensing cross-section. In addition to the criteria listed in \Sref{ssec:lenses}, we imposed a minimum magnification of $|\mu_{\mathrm{tot}}|=5$, for a system to be classified as a lens. Nine of the $\Ngold$ golden lenses have a measured magnification smaller than this threshold (nevertheless, we still included them in this new analysis). 
This new model provides a better match to the observed distribution in magnification and source position (\Fref{fig:betamu}), and the inferred model parameters are virtually identical to the fiducial case (\Fref{fig:cp}).
This is not surprising, because, for the purpose of inferring the mass distribution of the foreground galaxy population, the key property of the selection function model is the relative probability of galaxies of different mass to be strong lenses. Adding or removing a low-magnification tail to the sample changes the total number of lenses, but does not affect this property.
Nevertheless, the test suggests that the magnification may play a role in the detectability of a lens, giving us an important clue on how to improve our knowledge of the selection function.
However, given the low sensitivity of our measurements to the details of the lensing cross-section model, we do not attempt to refine it further, at this stage.

We also plan to improve our model for the distribution of elliptical galaxies in redshift--flux--colour space. The current model is sufficiently accurate for the purposes of this work, but is somewhat simplistic. 
As we argued in \Sref{sc:Photoz}, the limiting factor is the availability of spectroscopic redshifts for the lens galaxies, which we need to test the accuracy of our photo-$z$s.
With a larger pool of spectroscopic observations, both of lens and non-lens elliptical galaxies, we can increase the complexity of the model, possibly adopting a non-analytic approach, such as a self-organising map \citep[e.g.][]{Wri++20}. We can also include ground-based photometric observations to improve the precision of individual photo-$z$s.

\section{\label{sc:Concl} Conclusions}

We measured the total projected mass enclosed within an aperture of $\fivekpc$ of isolated elliptical galaxies, over the redshift range $0<z<1$. 
Although our measurement is based on strong lenses, which are a biased subset of the full galaxy population, we corrected for selection effects and constrained the properties of galaxies before lensing.
This is one of very few such measurements obtained so far, and the one that covers the broadest redshift range yet.

We used our measurement to compute the total mass-to-light ratio of elliptical galaxies as a function of their properties. We found that galaxies at $z=0.5$, with luminosity $\logten{(L_r\, h^2/L_{r,\odot})}=10.7$, and average size for their luminosity, have a ratio between their total mass within $\fivekpc$ and the total luminosity within the same aperture of $(6.7\pm0.3)\,M_\odot\,h/L_{r,\odot}$. 
We also measured how $\mfive$ varies with time. 
Although the exact evolutionary tracks of massive elliptical galaxies are uncertain, we can conclude that, in a scenario with a purely merger-driven growth and a passively evolving stellar population, a growth in stellar mass brings along a corresponding increase in $\mfive$.

According to our model, strong lens galaxies have, on average, a $10$--$20\%$ larger $\mfive$ compared to non-lens galaxies, with the difference being more pronounced at lower luminosity.
Since the precision of mass measurements from strong lensing is typically higher than this value, it is important to account for selection effects when generalising measurements made on strong lenses to the full population of galaxies.
This statement applies to studies based on statistical samples of lenses or individual lenses alike.

In summary, our measurement provides an observational benchmark with which to test models of galaxy structure, a unique one for the broad range of its redshift coverage, $0<z<1$. 
These measurements were obtained using data from the Q1 area of \Euclid, which constitutes less than $0.5\%$ of the total area that will be covered by the Euclid Wide Survey.
As the amount of data increases, we will use \Euclid strong lensing observations to investigate an increasing number of aspects of galaxy structure.

\begin{acknowledgements}
\AckEC
\AckQone
\AckDatalabs
AS acknowledges support from the National Key R\&D Program of China (No. 2023YFA1607800, 2023YFA1607802), and from the Yangyang Development Fund.
\end{acknowledgements}

%
%

\bibliography{Euclid,myreferences} 
%
%

%

\begin{appendix}
  
\section{Ambiguous lens candidates}\label{sect:ambiguous}

\begin{figure*}[htbp!]
  \begin{center}
    \includegraphics[width=\textwidth]{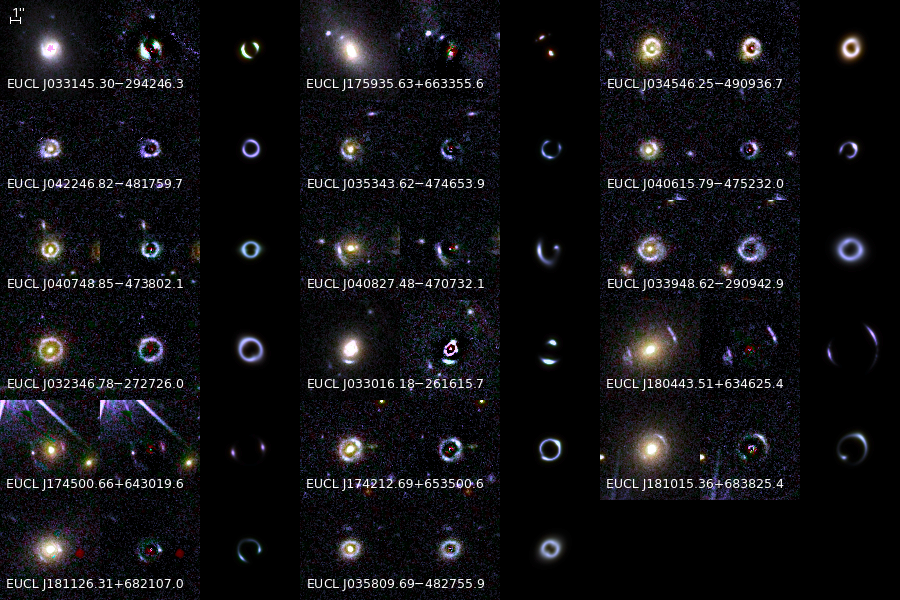}
   \end{center}
\caption{Colour-composite images of the $\Namb$ ambiguous lens candidates that we excluded from the golden lens sample. For each candidate, we show the corresponding lens light-subtracted image, and best-fit lens model next to it. Each image is $10''$ on the side.}
\label{fig:ambiguous}
\end{figure*}

As we explained in \Sref{ssec:lenses}, we adopted a series of objective criteria to define our sample of lenses. Nevertheless, we had to use judgement to further bring the purity of the sample to 100\%. In the process, we eliminated $\Namb$ ambiguous lens candidates that we deemed not to be lenses. Figure~\ref{fig:ambiguous} shows their images, alongside our best-fit model. In \Tref{tab:ambiguous} we list the motivations that led us to remove them from the golden lens sample.

\begin{table*}
\caption{
Ambiguous lens candidates excluded from the golden lens sample.
}
\smallskip
\label{tab:ambiguous}
\smallskip
\begin{tabular}{ll}
Object name & Reason for discarding \\
\hline
\hline
EUCL\,J033145.30$-$294246.3 & The model predicts an extended and curved counter-image, but the observed one is compact. \\
EUCL\,J175935.63+663355.6 & The counter-image is likely an object in the lens plane. \\
EUCL\,J034546.25$-$490936.7 & The ring shows irregular structure that is more typical of ring galaxies. \\
EUCL\,J042246.82$-$481759.7 & The shape of the arc cannot be reproduced by a lens model. \\
EUCL\,J035343.62$-$474653.9 & The model predicts an additional image that is not observed. \\
EUCL\,J040615.79$-$475232.0 & The model cannot reproduce the position of the counter-image. \\
EUCL\,J040748.85$-$473802.1 & The ring shows irregular structure that is more typical of ring galaxies.\\
EUCL\,J040827.48$-$470732.1 & The model requires a two-pixel offset between mass and light to reproduce the images. \\
EUCL\,J033948.62$-$290942.9 & The surface brightness of the ring is too extended, compared to that predicted by the model. \\
EUCL\,J032346.78$-$272726.0 & The ring shows irregular structure that is more typical of ring galaxies. \\
EUCL\,J033016.18$-$261615.7 & The lens shows disturbed morphology, making a non-lens scenario more likely. \\
EUCL\,J180443.51+634625.4 & The arcs are too straight, compared to the model prediction. \\
EUCL\,J174500.66+643019.6 & The relative brightness of the two images is very different from that predicted by the model. \\
EUCL\,J174212.69+653500.6 & The ring shows irregular structure that is more typical of ring galaxies. \\
EUCL\,J181015.36+683825.4 & The shape of the counter-image does not match that predicted by the model. \\
EUCL\,J181126.31+682107.0 & The orientation of the arc is different from that predicted by the model. \\
EUCL\,J035809.69$-$482755.9 & The ring shows irregular structure that is more typical of ring galaxies. \\
\end{tabular}
\end{table*}

\section{\label{sect:compl} Completeness}

Our analysis rests on the assumption that our lens sample is a representative subset of all of the golden lenses in Q1. That is, we assume that our sample is either complete or is missing lenses at random, with no correlation between the physical properties of the lenses and the probability of them being included in the sample.
Assessing completeness is difficult because we lack a truly complete reference sample of lenses, in Q1 or anywhere.
Our main approach is to examine how the lens finding procedure that fed our sample performed on 1) images of simulated lenses, and 2) lenses that were found directly from visual inspection.
The first step in this process is to make a census of the provenance of the golden lenses.
Of the $\Ngold$ lenses in our sample, $\NSLDE$ were discovered as part of the search described by \SLDEA. Two of them were previously known: EUCL\,J041339.51$-$484111.4 was originally discovered by \citet{Sto++24}, and EUCL\,J040457.32$-$492221.5 was in the sample of \citet{Inc++26}. 
Of the remaining 16 golden lenses, eleven were discovered by \citet{SLDEF}. These lenses were excluded by \SLDEA\ at the sample-building phase, as they removed any object with a match in the \emph{Gaia} catalogue, as an attempt to remove stars from the sample. 
Two lenses were discovered by \citet{SLDEB}, one by \citet{Q1-Xu}, and one, EUCL\,J040205.11-480808.4, was a previously known lens, discovered by \citet{Ste++22}.
Finally, our sample includes the lens NGC\,6505, which was originally discovered by \citet{ORiordan25}, and which would have been found by the \SLDEA\ search as well.
In summary, the vast majority of our lenses belong to the \SLDEA\ sample. Therefore, we focus primarily on that lens finding effort to assess the completeness of our sample.

\SLDEA\ estimated the completeness of their lens search to be in the range $60$--$70\%$. However, that figure refers to all candidates graded B or above. As the golden lenses are a much smaller subset, we expect the completeness of the golden lens sample to be different.
The lens search of \SLDEA\ consisted of a three-step process. In a first step, potential lens candidates were ranked by making use of convolutional neural networks (CNNs). In the second step, the highest-scoring \num{78214} lens candidates were visually inspected by citizen scientist volunteers, along with $40\,000$ candidates drawn randomly from the initial pool of $\sim10^6$ objects (these two samples overlap). In the third and final step, expert scientists visually inspected $7362$ lens candidates and assigned them the grades on which our initial selection is based.

\begin{figure}
\includegraphics[width=\columnwidth]{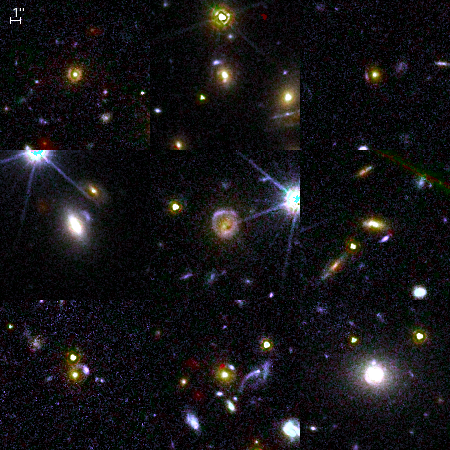}
\caption{
Selection of simulated golden lenses that were classified as grade C or below by lensing experts.
\label{fig:expert_missed}
}
\end{figure}

We can evaluate the completeness of the human selection steps by analysing their performance on images of simulated lenses.
These simulations were shown to experts and volunteers during the classification process, interspersed among the real images of lens candidates.
Experts collectively classified 265 simulated lenses, $146$ of which are golden lenses. Of these, $114$ were given a grade B or above, and $32$ (approximately $22\%$) were missed. The images of a subset of those lenses are shown in \Fref{fig:expert_missed}.
 By comparison, volunteers reached a much higher completeness. Of $1861$ simulated golden lenses classified by them, only $35$ were rejected (less than $2\%$). This is not surprising, because the goal of experts is to strike a balance between completeness and purity, and the purity of their selection is much higher than that of the volunteers.
Taking these numbers at face value, one might conclude that the completeness of the two human inspection steps combined is roughly $77\%$, with the limiting factor being the expert classification.
However, caution is required. This figure is based on the implicit assumption that the simulated golden lenses are representative of the real sample of golden lenses in the sky. We have no means of verifying this assumption at this stage. 
In order to make progress in our understanding of the human selection function, classification data on a much larger sample of simulated golden lenses are needed. We aim to build such a data set in future lens searches.

To assess the completeness of the CNN-based selection step, we examined the random sample of images that was independently inspected by the volunteers.
In the \num{40000} randomly selected objects lie six golden lenses. Five of these were also flagged by the CNNs, while one was missed. 
Therefore, our best guess for the completeness of the CNN selection step is $\sim83\%$. However, this estimate is highly uncertain due to the small numbers involved. 

Multiplying the completeness of the CNN- and visual-selection steps that we estimated above, we obtain a value of $64\%$ for the completeness of the \SLDEA\ lens finding procedure. On the one hand, this is a lower limit of the overall completeness of our sample, because some of the lenses missed by \SLDEA\ were later found by \citet{SLDEB}, \citet{Ber++26}, \citet{SLDEF}, and \citet{Q1-Xu}. On the other hand, the limited amount of information available prevents us from drawing a firm conclusion on completeness at this stage.
With future lens searches, we aim to both increase our completeness over the Q1 area and to better characterise the selection function of our search.
\end{appendix}

\label{LastPage}
\end{document}